\documentclass{article}
\title{Evolutionary shift detection with ensemble variable selection}
\usepackage[utf8]{inputenc}
\usepackage{graphicx}
\usepackage{caption}
\usepackage{setspace} 
\usepackage{lineno}
\usepackage{hyperref}
\usepackage[margin=1in]{geometry}

\let\oldequation\equation
\let\oldendequation\endequation

\renewenvironment{equation}
  {\linenomathNonumbers\oldequation}
  {\oldendequation\endlinenomath}

\usepackage{xcolor}  
\usepackage{amsmath}
\DeclareMathOperator{\loglik}{loglik}
\DeclareMathOperator{\BIC}{BIC}
\DeclareMathOperator{\pBIC}{pBIC}
\DeclareMathOperator{\pa}{pa}
\DeclareMathOperator{\pathm}{path}
\DeclareMathOperator{\nodeend}{end}
\usepackage[round]{natbib}
\providecommand{\keywords}[1]
{
  \small	
  \textbf{\textit{Keywords---}} #1
}
\usepackage{fancyhdr}
\author{%
Wensha Zhang$^{1}$, Toby Kenney$^{2*}$, Lam Si Tung Ho$^{3*}$
}
\begin{document}
\begin{titlepage}
\maketitle

\begin{center}
{\footnotesize 
$^{1,2,3}$Department of Mathematics and Statistics, Dalhousie University, Nova Scotia, Canada\\
$^{*}$These authors have equal contribution\\
E-mails: \{wn209685,tb432381, Lam.Ho\}@dal.ca
}
\end{center}

\begin{abstract}
\noindent
\begin{enumerate}
\item Abrupt environmental changes can lead to evolutionary shifts in trait evolution.
Identifying these shifts is an important step in understanding the evolutionary history of phenotypes.

\item We propose an ensemble variable selection method (\texttt{R}
  package ELPASO) for the evolutionary shift detection task and
  compare it with existing methods (\texttt{R} packages $\ell$1ou and
  PhylogeneticEM) under several scenarios.
  
\item The performances of methods are highly dependent on the selection
  criterion. When the signal sizes are small, the methods using the
  Bayesian information criterion (BIC) have better performances. And
  when the signal sizes are large enough, the methods using the
  phylogenetic Bayesian information criterion (pBIC)
  \citep{khabbazian_kriebel_rohe_ane_2016} have better
  performance. 
  Moreover, the performance is heavily impacted by 
measurement error 
and tree reconstruction error. 

\item Ensemble method + pBIC tends to perform less conservatively than $\ell$1ou + pBIC, and Ensemble method + BIC is more conservatively than $\ell$1ou + BIC. PhylogeneticEM is even more conservative with small signal sizes and falls between $\ell$1ou + pBIC and Ensemble method + BIC with large signal sizes. The results can differ between the methods, but none
  clearly outperforms the others. By applying multiple methods to a
  single dataset, we can access the robustness of each detected shift, based on the agreement among methods.

\end{enumerate}
\end{abstract}


\end{titlepage}

\section{Introduction}
Understanding the evolutionary process of species is an important task
in phylogenetic comparative studies. \citet{felsenstein_1985} is the
first to introduce Brownian Motion (BM) to model the evolution of a continuous trait. 
BM models have been used in many evolutionary studies, such as flower size
evolution (\citealp*{davis_latvis_nickrent_wurdack_baum_2007}), genome size evolution \citep{beaulieu2012modeling}, the
spread of HIV-1 in central Africa
(\citealp*{gill_tung_ho_baele_lemey_suchard_2016}), and mammalian life
history traits
(\citealp*{hassler_tolkoff_allen_ho_lemey_suchard_2020}). \citet{hansen_1997}
proposed to use an Ornstein-Uhlenbeck (OU) processes to model evolution with natural selection. Unlike the
BM process, the variance of the OU process in traits is bounded, which is
more realistic (\citealp*{butler_king_2004}). Therefore, OU processes
are now widely used in many evolutionary studies, including character
displacement in Lesser Antillean Anolis Lizards
(\citealp*{butler_king_2004}), and HIV-1 heritability
(\citealp*{bastide_ho_baele_lemey_suchard_2021}).

\citet{butler_king_2004} formulate the multiple optima OU model for
adaptive evolution in which optima differ between branches, and remain
constant along an evolutionary path until discrete events where
changes in selective regime occur. They use hypothesis testing to
test whether the optima are different between groups. The changes can
be modeled as shifts in the parameters of the OU processes. The shifts in
optima are considered to be correlated with abrupt environmental
changes (\citealp*{losos_2011};
\citealp*{mahler_ingram_revell_losos_2013}). For example,
\citet{jaffe_slater_alfaro_2011} investigate the relationship between
the difference in optimal body sizes and habitat changes for
turtles. Therefore, by detecting shifts in optima based on observed
traits, we can get knowledge of unobserved historical environmental
changes and better understand the evolutionary process of species. Our
goal here is to use statistical models to detect the evolutionary
shifts: where the shifts have occurred and the size of the
shifts.

There are some existing approaches to address this
problem. \citet{uyeda_harmon_2014} propose a Bayesian framework to
detect shifts in the selective optimum of OU models. However, results
of Bayesian approaches are deeply influenced by prior distributions
and the computation cost is relatively high.  We will focus on
frequentist approaches in this paper.  \citet{ho_ane_2014} illustrate
the limitation of traditional model selection criteria (AIC, BIC) in
the shift detection task and propose to use forward-backward
selection with modified BIC (\citealp*{zhang_siegmund_2006}).
\citet{khabbazian_kriebel_rohe_ane_2016} formulate
the shift detection problem into a variable selection problem and
combine the OU model with LASSO to detect the shift points, which
they implement in the $\ell$1ou \texttt{R} package. 
\citet{bastide_mariadassou_robin_2016} develop a maximum likelihood
estimation procedure based on the EM algorithm (implemented in the \texttt{R} package PhylogeneticEM).

Recently, ensemble methods have been widely applied to variable
selection problems (\citealp*{lee_leu_2011};
\citealp*{piao_piao_park_ryu_2012};
\citealp*{mera-gaona_lopez_vargas-canas_2021}). Ensemble feature
selection can add more diversity to selected variables and produce
robust variable selection results (\citealp*{pes_2019}).
In this paper, we propose an ensemble variable selection method for
shift detection and compare it with existing methods implemented in the
PhylogeneticEM and $\ell$1ou packages. We have implemented our method
in a new \texttt{R} package, \texttt{ELPASO} (Ensemble LASSO for Phylogenetic Analysis of Shifts with OU). It is available at \url{https://github.com/WenshaZ/ELPASO}.

The following sections are organized as follows. Section \ref{sec2} introduces
the problem formulation of the shift detection task as a variable
selection problem, and presents the methodology of $\ell$1ou,
PhylogeneticEM and our ensemble method.  Section \ref{sec3} presents our comparison results on simulated data under OU models. In order to better evaluate the performance, we use three different
criteria: true positive and false positive shifts detected, predictive log-likelihood on a test data set, and adjusted rand index. Section \ref{sec4}
discusses the effect of shift position and tree shape on the performances of detection methods. 
Section \ref{sec5} compares the methods when the model assumptions are not satisfied.
Previous simulation studies often ignore this scenario. However, in practice, the
model assumptions are usually violated. 
Section \ref{real_data} presents the case study results on Anolis Lizard data. 
Finally, Section \ref{sec7} provides conclusions and discussion.

\section{Shift detection for trait evolution models} \label{sec2}

\subsection{Trait evolution models}
The trait evolution process on a phylogenetic tree describes the
changes in traits through generations. Each species has a certain
value of the trait. And the trait values of different species are
correlated because of their shared ancestry which is represented by a phylogenetic tree. We only observe the trait values at the tips of the tree. 
In this paper, the tree is assumed to be separately estimated from
sequence data, and is treated as known.

Trait evolution models are used to model
how the trait values change over time. Brownian Motion and
Ornstein–Uhlenbeck are two commonly used models to model the evolution of continuous traits. 
Let $\mathrm{Y}$ denote the vector of observed trait values
at the tips, $Y_i$ as the trait value of taxon $i$. 
These two models assume that conditioning on the trait value of a parent, the evolutionary processes of sister species are independent. 
So we only need to specify the model on one branch. For a single branch, we let $Y(t)$ denote the trait value at time $t$.

\subsubsection*{Brownian Motion Model}
\citet{felsenstein_1985} proposed to use Brownian motion (BM) to model
the evolution of continuous traits over time.  Trait values of sister
lineages start at the trait value of their most recent common ancestor
and evolve independently following a BM model. The result of this
model is that the correlation between the trait values of two species
depends only on the evolution time they shared.  Under this model, the
observed trait values $\mathrm{Y}$ follow a multivariate Gaussian
distribution. For an ultrametric tree of height $1$, each $y_i$ has mean
$\mu_0$ and variance $\sigma^2$, and the covariance between $y_i$ and
$y_j$ is $\sigma^2 t_{ij}$, where $t_{ij}$ is the shared evolution
time between species $i$ and $j$.

\subsubsection*{Ornstein–Uhlenbeck Model}

The variance of the BM model is unbounded, which is considered
unrealistic. 
The OU model \citep{hansen_1997}, on the other hand, incorporates a selection force that pulls the trait value toward a selective optimum $\theta$. This model is preferable to the BM model because of its more realistic assumptions. 
An OU process $Y(t)$ is defined by the following stochastic differential equation
\begin{linenomath*}
\begin{equation} dY(t)=\alpha[\theta(t)-Y(t)]dt+\sigma dB(t), \end{equation}
\end{linenomath*}
where $ dY(t)$ is the infinitesimal change in trait value; $B(t)$ is a standard BM; $\sigma^2$ measures the intensity of random fluctuation; $\theta(t)$ is the optimal value of the trait at time $t$; and $ \alpha \geq 0$ is the selection strength. When $\alpha = 0$, the OU process is the same as a BM. We assume that $\alpha$ and $\sigma$ are constant.

\subsection{Shift detection as a linear model selection problem}
For the OU model, the assumption that the optimal value $\theta(t)$ is
constant throughout the tree is not realistic, as different trait
values are suited for different environments and evolutionary
strategies. A more practical model allows the optimal value
$\theta(t)$ to shift at certain positions on the
tree. 
\citet{hansen_1997} proposed a heterogenous OU model to allow different optimal values on different branches. 
Shifts are noncontinuous changes in the optimal value during the evolution process. 
A shift on a branch of the phylogenetic tree would influence
all the species under that branch. Our goal here is to find the
positions of shifts and estimate the changes in optimal trait value,
$\theta$, at the shifts. We assume that the optimal value on one
branch is constant. Let $\theta_b$ denote the optimal value on branch
$b$ and $t_b$ the age of the beginning of branch $b$. Let $T$ denote
the age of the root node. We only consider ultrametric trees in this shift detection task.
Let $\pa(b)$ denote the parent edge of $b$
and $\nodeend(b)$ the end node of $b$. Thus, $\triangle \theta_b =
\theta_{\pa(b)}-\theta_b \neq 0$ means that a shift in optimal value
occurred on branch $b$.

Let $\mathbf{Y}$ denote the observed trait values at the tips, $Y_i$
denote the trait value of taxon $i$ and $Y_0$ denote the trait value of
the root node. Under the OU process, $\mathbf{Y}$ follows a
multivariate normal distribution. The mean of each random variable
$Y_i$ is (\citealp*{hansen_1997}):

\begin{equation}
   E(Y_i) = Y_0e^{-\alpha T}+\sum_{b \in \pathm(\textrm{root},i)}\left(e^{-\alpha t_{\nodeend(b)}}-e^{-\alpha t_b}\right)\theta_b 
\end{equation}

The covariance between $Y_i$ and $Y_j$ is \citep{ho2013asymptotic}:

\begin{equation}
\label{eqn:covariance}
    \Sigma^{(\alpha)}_{ij}=\left\{
\begin{array}{rcl}
\sigma^2e^{-\alpha d_{ij}}(1-e^{-2\alpha t_{ij}})/(2\alpha)       &      & \textrm{fixed root model}\\
\sigma^2e^{-\alpha d_{ij}}/(2\alpha)     &      & \textrm{random root model}
\end{array} \right. 
\end{equation} 
where $t_{ij}$ is the shared time between species $i$ and $j$, and $d_{ij}$ is the
distance between $i$ and $j$. 
In the random root model, the trait value at the root is assumed to follow the stationary distribution.
To transfer the shift detection problem into
a regression problem, we can rewrite the mean of $Y_i$ as \citep{khabbazian_kriebel_rohe_ane_2016}:

\begin{equation}
    \begin{aligned}
\label{eqn:trait-mean}
    E(Y_i) &= Y_0e^{-\alpha T}+\sum_{b \in \pathm(\textrm{root},i)}\left(e^{-\alpha t_{\nodeend(b)}}-e^{-\alpha t_b}\right)\theta_0\\
    &+\sum_{b \in \pathm(\textrm{root},i)} \sum_{b' \in \pathm(\textrm{root},b)}\left(e^{-\alpha t_{\nodeend(b)}}-e^{-\alpha t_b}\right)\triangle\theta_{b'}\\
    &=Y_0e^{-\alpha T} + (1-e^{-\alpha T})\theta_0 + \sum_{b' \in \pathm(\textrm{root},i)} \sum_{b \in \pathm(b',i)}\left(e^{-\alpha t_{\nodeend(b)}}-e^{-\alpha t_b}\right)\triangle\theta_{b'} \\
    &= Y_0e^{-\alpha T} + (1-e^{-\alpha T})\theta_0 + \sum_{b' \in \pathm(\textrm{root},i)}(1-e^{-\alpha t_{b'}})\triangle\theta_{b'}
    \end{aligned}
\end{equation}
Let $\beta_0 = Y_0e^{-\alpha T} + (1-e^{-\alpha T})\theta_0$ and $\beta_b = (1-e^{-\alpha t_{b}})\triangle\theta_{b}$. In this way, the shift detection problem under the OU model can be converted to a linear model selection problem. The trait values at tips can be written as:

\begin{equation}
   \mathbf{Y} =\beta_0\mathbf{1}+\sum_b\beta_b \mathbf{X_b}+\mathbf{\epsilon} 
\end{equation}

where $X_b$ is a vector defined by $X_{bi}=0$ if taxon $i$ is not under
branch $b$, and $X_{bi}=1$ if the taxon $i$ is under branch $b$, and
$\epsilon$ follows a normal distribution with mean 0 and covariance
matrix $\Sigma^{(\alpha)}$, which is given by Equation~\ref{eqn:covariance}
The main task is to select the branches that have $\beta_b \neq 0$.

\subsection{$\ell$1ou}
\citet{khabbazian_kriebel_rohe_ane_2016} propose a phylogenetic LASSO
method to detect shifts in optimal trait value under OU models.  To remove the
influence of the covariance matrix, they conduct a transformation
before model selection:

\begin{equation}
\Sigma_{\alpha}^{-1/2}\mathbf{Y} =\beta_0\Sigma_{\alpha}^{-1/2}\mathbf{1}+\Sigma_{\alpha}^{-1/2}\mathbf{X}\mathbf{\beta}+\Sigma_{\alpha}^{-1/2}\mathbf{\epsilon}
\end{equation}

Where $\mathbf{\beta}$ denotes the vector of $\beta_b$ and $\mathbf{X}$ denotes the design matrix, the $b\text{th}$ column of $\mathbf{X}$ is $\mathbf{X_b}$. After data transformation, the error terms $\Sigma_{\alpha}^{-1/2}\epsilon$ become a vector of independent standard normal random variables. The LASSO solution is to minimize the least squares with $\ell 1$ penalty. The loss function is given by

\begin{equation}
    \frac{1}{2}\|\Sigma_{\alpha}^{-1/2}\mathbf{Y}-\beta_0\Sigma_{\alpha}^{-1/2}\mathbf{1}-\Sigma_{\alpha}^{-1/2}\mathbf{X}\mathbf{\beta}\|^2+\lambda\|\beta\|_1
\end{equation}

They use the package \texttt{lars} to estimate $\beta$ for every
$\lambda$ value, and conduct backward selection based on the model
selection criterion pBIC (see Section 2.6 for more details about pBIC)
using the models selected for each $\lambda$ value as a starting
point. They then use the same criterion to select from among the
models found for different values of $\lambda$.

The above process is based on a given $\alpha$ value.
\citet{khabbazian_kriebel_rohe_ane_2016} use the following procedure to obtain the estimation of $\alpha$. Firstly, they set $\alpha=0$ and run the
variable selection for this value of $\alpha$. They then refit
$\alpha$ with the selected variables. They repeat the selection step
for the new $\alpha$, and choose the model with the best criterion score
from among these models. Because of the use of LASSO, the computation
speed is faster than previous shift detection tools, including SURFACE (\citealp*{ingram_mahler_2013}) and bayou (\citealp*{uyeda_harmon_2014}).
Their implementation of the method is available in the \texttt{R} package $\ell$1ou.

\subsection{PhylogeneticEM} \label{phyloEM}
\citet{bastide_ane_robin_mariadassou_2018} introduce a framework which
treats phylogenetic analysis as a missing data problem, allowing the
usage of the EM algorithm.  They set $\mathbf{\tau} = (\sigma,\alpha,
\mathbf{\beta})$ as the vector of all the parameters to estimate, and
$\textbf{X = (Z,Y)}$ as the trait values of both internal and external
nodes.
$\textbf{Z}$ is
the vector of trait values of internal nodes, $\textbf{Y}$ is the
vector of trait values of external nodes.

They assume the number of shifts is fixed and use the EM algorithm to
estimate the parameters by maximizing the log likelihood
$\log{p_\tau(\mathbf{Y})}$.  The EM algorithm is based on the
decomposition:
\begin{equation}
    \log{p_\tau(\mathbf{Y})} = E_\tau[\log{p_\tau(\mathbf{Z,Y})}|\tau]-E_\tau[\log{p_\tau(\mathbf{Z|Y})}|\mathbf{Y}]
\end{equation}
The difficulty with the maximization, in this case, comes from the fact
that the locations of shifts on the branches are discrete
variables. They used a Generalized EM (GEM,
\citet{dempster_laird_rubin_1977}) to conduct the maximization. The
complexity for this is $O(n^k)$ where $k$ is the number of shifts.

The above process is based on the assumption that the number of shifts
$k$ is fixed. They estimate the parameters with $k = 1, ...., K$,
where $K$ is the given maximum number of shifts. They then conduct
model selection on $k$ based on penalized least squares:  
\[
\text{PLS} =
\left(1+\frac{\text{pen}(k)}{n-k-1}\right)\sum_{i=1}^p\|Y_i-\hat{Y_i}\|^2
\]
where $\hat Y_i$ is the predicted trait value of taxon $i$ given by the
model with $k$ shifts. The penalty term $\textrm{pen}(k)$ if given by
\[
\text{pen}(k) =
A\frac{n-k-1}{n-k-2}\text{EDkhi}\left[k,n-k-2,\frac{(k+1)^2}{|S_k^{PI}|}\right]
\]
where $|S_k^{PI}|$ denotes the number of parsimonious identifiable sets of
locations of k shifts and $A$ is a constant which the authors fixed at
1.1 based on simulation results.  The EDkhi function
(\citealp*{baraud_giraud_huet_2009}) is defined as follows:
\[
\textrm{DKhi}(D,N,x)=
\frac{1}{\text{E}(X_D)}\text{E}\left[\left(X_D-x\frac{X_N}{N}\right)_+\right]
\]
where $D$ and $N$ are positive integers, $X_D$ and $X_N$ are independent chi-squared random variables with
$D$ and $N$ degrees of freedom respectively.  Then, for $0<q\leq1 $,
$\textrm{EDkhi}(D,N,q)$ is defined as the unique solution to
\[
\text{Dkhi}[D,N,\text{EDkhi}[D,N,x]] = q.
\]According to
\citet{baraud_giraud_huet_2009}, the penalty used here bounds the risk
of the selected variables, and gives non-asymptotic guarantees. The
implementation of this method is available in the \texttt{R}
package \texttt{PhylogeneticEM}.

\subsection{Ensemble method}
The framework of the ensemble variable selection model for shift
detection consists of two phases. Firstly, we apply LASSO on a number of random subsamples of the data. 
For each subsample, we obtain
a ranking of the variables based on the largest penalty $\lambda$ for
which the variable is selected by LASSO. We aggregate the rankings
from each subsample into an overall variable ranking. Secondly, we
use this ranking as a basis for a variable selection method.

\begin{figure}
  \centering
    \includegraphics{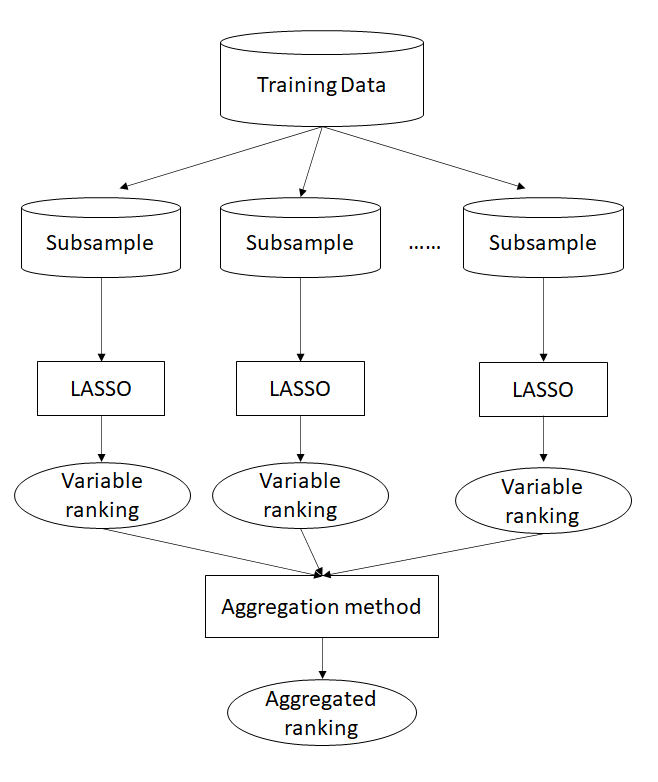}
     \caption{\label{Fig1}The model structure of ensemble method for shift detection}
\end{figure}

The foundation of ensemble learning is to combine the results of
multiple models. The idea is that combining the results of several
models will obtain better results by reducing the model variance and
bias. Bagging and boosting are the two most commonly used ensemble
models. There has been substantial recent work on the use of ensemble
models for feature selection.
\citet{bolon-canedo_alonso-betanzos_2018} summarize the different
types of ensemble methods that are used in feature selection. We here
use a homogenous scheme for the ensemble. That is, we firstly take
subsamples from the training dataset, then apply LASSO to each
subsample. LASSO provides a solution path by varying the penalty
size. Therefore, for each subsample, a variable ranking is produced
where variables are ranked in decreasing order of the largest penalty
for which they are selected. Aggregating the ranking sequences from
all the subsamples, we can get the overall ranking for all the
variables. The process is shown in Figure~\ref{Fig1}.

There are several choices for how we aggregate the rankings from the
different subsamples into a single overall ranking. For example,
geometric mean, arithmetic mean, median, and so on. It is possible to apply
these aggregation methods either to the ranking or to the values of
the penalty $\lambda$. We suggest using the first quantile of the
ranking to aggregate the results because the first quantile is robust
to outliers. In a few subsamples, a shift may be ranked very low,
possibly because the taxa that distinguish it from other shift
positions are not included in the subsample. In these cases, the rank
might be very large, and therefore have a large influence on the
geometric or arithmetic mean. Similarly, the first quantile help improving our ability to distinguish between shifts and surrogates. It is common for only one of the
surrogate variables to be ranked highly, and the shift to have a low
rank. This can cause the median rank of the true shifts to be low.

After obtaining the aggregated ranking, we use model selection to
choose the final combination of variables. Forward selection, backward
selection, and stepwise selection are potential approaches.
Forward selection starts with the null model, and sequentially adds
variables in the ranked order, starting with the highest ranked, as
long as the model score is better than the previous model. Backward
selection starts with the full model, and sequentially removes
variables in rank order, starting with the lowest ranked, as long as
the model score after removing each variable is better than the
previous model. Stepwise selection consists of one forward
selection pass followed by one backward selection pass, starting with
the model selected by forward selection. Firstly include the variables
that improve the model score then remove the variables whose removal
further improves the model score. From the simulation results,
stepwise selection performs best.

Our procedure to estimate $\alpha$ and $\sigma^2$ is similar to
$\ell$1ou.
\begin{enumerate}
    \item Fit a null BM phylogenetic regression model on the dataset. Get the initial estimate of $\sigma^2$.
    \item Use $\alpha=0$ and the $\sigma^2$ value from the first step to calculate the covariance matrix and apply the ensemble method selection procedure.
    \item Fit the phylogenetic regression model 
      with the selected variables in Step~2 and
      get new estimates of $\alpha$ and $\sigma^2$.
    \item Repeat Step~2 and Step~3 once more.
    \item Select the model with the best model criterion score
\end{enumerate}

\subsection{Model selection criteria}
For all the models, a criterion is used to conduct model selection for the number of shifts. 
AIC and BIC are most the commonly used criteria in model selection
problems. \citet{ho_ane_2014} showed that using AIC as the criterion
may lead to model
overfitting. \citet{khabbazian_kriebel_rohe_ane_2016} present a new
criterion pBIC including a phylogenetic correction. The traditional
BIC is given by:
\[
\BIC(M_k) = -2\loglik(M_k)+(2k+3)\log(n)
\]
where $n$ is the number of taxa, $k$ is the number of shifts selected, $M_k$ is the estimated model. $2k+3$ is the number of parameters: each shift location and magnitude is counted as a parameter and there are 3 general parameters ($\beta_0$, $\alpha$, and $\sigma$).
The phylogenetic BIC proposed by \citet{khabbazian_kriebel_rohe_ane_2016} is given by:
$$\pBIC(M_k) = -2\loglik(M_k)+2k\log(2n-3)+2\log(n)+\log
\text{det}((X_{M_k}^{\hat\alpha})^{T}v
\Sigma_{\alpha}^{-1}X_{M_k}^{\hat\alpha}) $$
where
$X_{M_k}^{\hat\alpha}$ is the matrix $X^{\alpha}$ with only the
columns corresponding to the k selected shifts, and $v$ is the
observed trait variance. 
The penalty for the shift position is $2k\log(2n-3)$. 
The penalty for shift magnitudes and the intercept are shown in the last term.
PhylogeneticEM uses penalized least squares for model selection;
details are in Section~\ref{phyloEM}.

\section{Simulation studies}
\label{sec3}

We conduct simulations to compare PhylogeneticEM, $\ell$1ou (pBIC/BIC)
and ensemble LASSO (pBIC/BIC). The most direct method for comparison is to compare how many true shifts the methods detect and
how many wrong shifts that are selected by them. However, the OU model is not completely identifiable
\citep{bastide_mariadassou_robin_2016, ho_ane_2014, khabbazian_kriebel_rohe_ane_2016}, and even if the selected
shifts are not equivalent to the true model, a good argument can be
made that selecting a close surrogate shift is preferable to failing
to select the shift at all. In these cases, the true positive versus false positive curve might misrepresent the performance, since neither
method has a true positive, but the method that selects the surrogate
has a false positive, and so is deemed to have performed worse, even
though selecting the close surrogate is arguably more
correct. Therefore, we include two more measurements in comparison:
predictive log-likelihood and adjusted rand index. The idea of
predictive log-likelihood is to compare the prediction accuracy on
test data, of the selected models from different methods. Adjusted
rand index evaluates how similar the clustering of the selected
model is to the clustering of the true model. We can get a more
comprehensive understanding of the characteristics, strengths,
limitations of the methods by combining the three different
measurements.

\begin{figure}
  \centering
    \includegraphics[width=\textwidth]{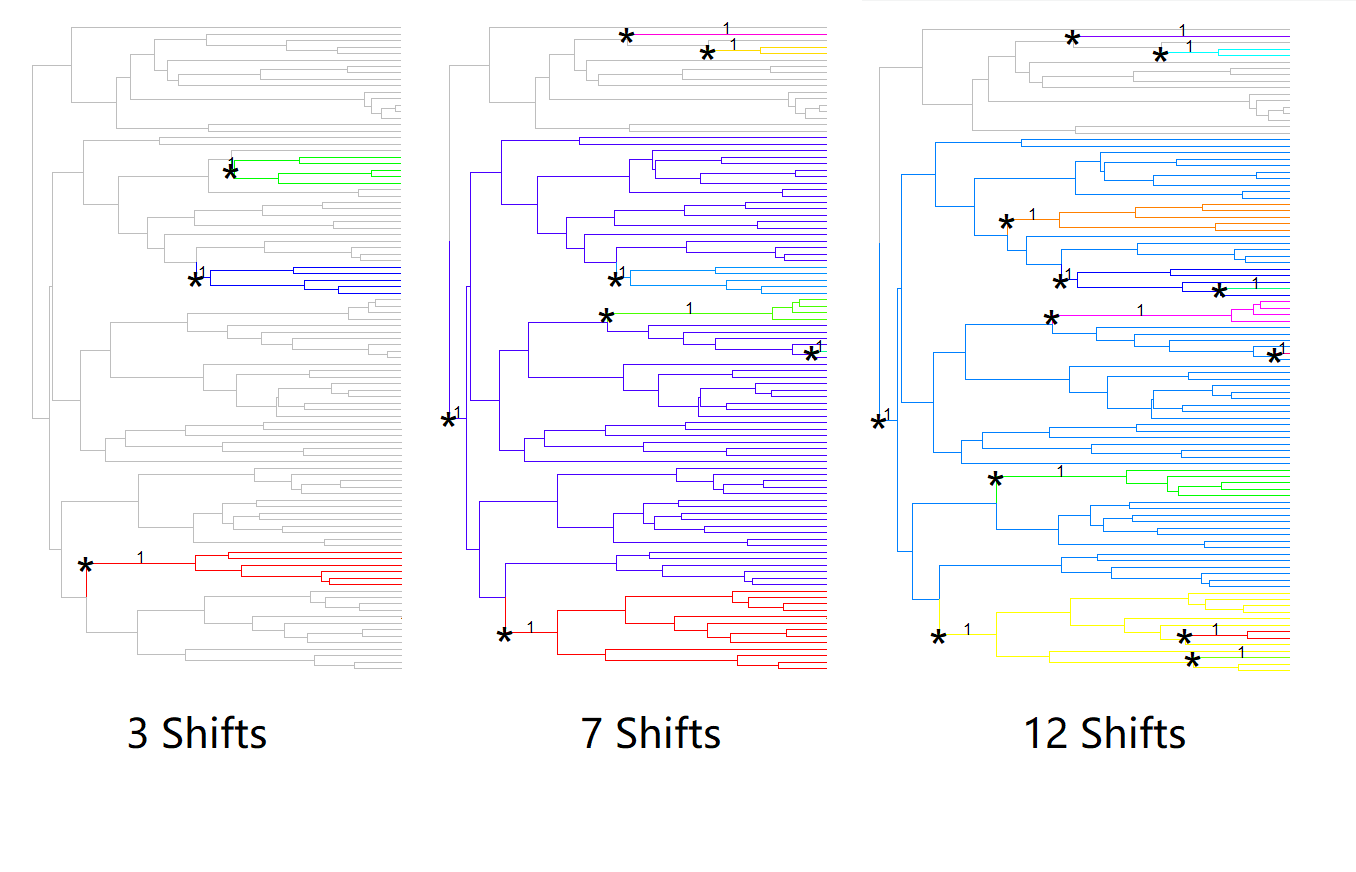}
     \caption{\label{Fig2}Tree used in simulations to compare the precision and recall of different methods. The shifts positions are indicated by asterisks. Different colours indicate different optimal values for the trait.}
\end{figure}

We simulate a number of scenarios with varying numbers of shifts and
signal sizes. We simulate datasets under OU models along the 100-taxon
Anolis lizards’ tree (\citealp*{mahler_ingram_revell_losos_2013}). We
compare the methods on scenarios with 3, 7, or 12 shifts. Data were
simulated according to the shifts in Figure~\ref{Fig2}. For each scenario, we
set $\alpha = 1$ and $\sigma^2 = 2$ so that $\frac{\sigma^2}{2\alpha}
= 1$. We simulate under seven true signal sizes: $\beta=0.2, 1, 1.5, 2, 2.5, 3$ and
$5$. In each simulation, all shifts have the same value of $\beta$.

\subsection{True positive versus false positive}

We first compare the methods by true positive versus false positive
curves. True positive is the number of true shifts detected by the
model. False positive is the number of shifts that are not simulated
but which are wrongly detected by the model. If two models have similar false
positive values, the one with a higher true positive value is
considered to have a better performance. If one model has a higher true
positive and higher false positive than another, there is no
obvious conclusion about which model is better. It becomes a trade-off
problem between precision and recall. 

\begin{figure}
  \centering
    \includegraphics[width=\textwidth]{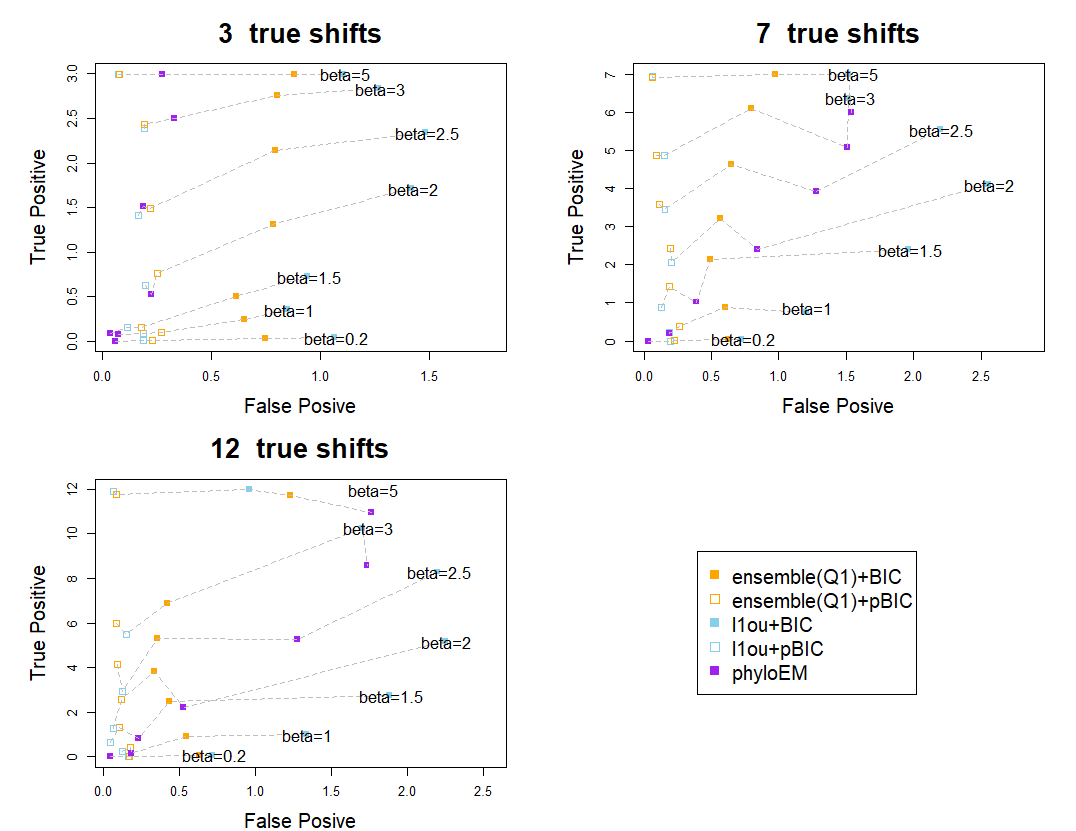}
 \caption{\label{Fig3}True positive numbers versus False positive numbers with
   3,7,12 shifts.}
\end{figure}

Figure~\ref{Fig3} shows the average true positive versus false
positive curve from 200 simulations in each scenario. Each point in
Figure~\ref{Fig3} represents the mean of true positive and mean of
false positive values. From the simulation results, $\ell$1ou+pBIC is
usually the most conservative method, with both lowest true positive
and false positive. For example, in the simulation of 7 shifts and
$\beta=2$, $\ell$1ou+pBIC on average detects under 2
shifts. $\ell$1ou+BIC is usually the least conservative
method. Ensemble LASSO provides more balanced choices between those
two methods. In most simulations, ensemble LASSO with BIC and pBIC
have both higher true positive and false positive compared to
$\ell$1ou+pBIC and have both lower true positive and false positive
compared to $\ell$1ou+BIC. Furthermore, in some situations, ensemble
LASSO methods have a better performance compared to $\ell$1ou. For
example, in the simulation of 7 shifts and $\beta=2$, ensemble
LASSO+pBIC has higher true positive and lower false positive compared
to $\ell$1ou+pBIC and ensemble LASSO+BIC has higher true positive and
lower false positive compared to $\ell$1ou+BIC. PhylogeneticEM is even
more conservative than $\ell$1ou+pBIC when the signal sizes are
small. It performs well where there are only 3 true
shifts. However when the number of shifts and the coefficient sizes
are large enough, PhylogeneticEM performs poorly compared to other
methods, including more false positive and fewer true positive
variables.

\subsection{Predictive log-likelihood}

\begin{figure}

  \centering
    \includegraphics[width=\textwidth]{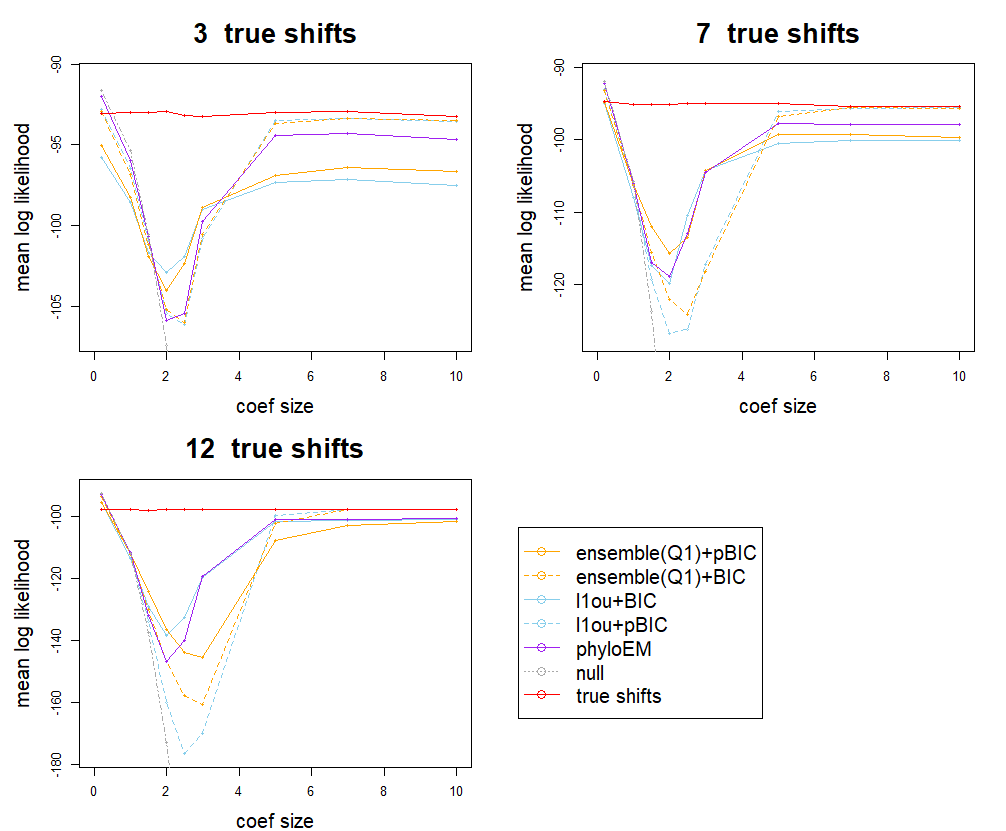}
     \caption{\label{Fig4}The mean log likelihood on 1000 test datasets}
\end{figure}

For each simulation, we generate $1000$ test datasets and $200$ training
datasets. 
We calculate the mean of log likelihood values over $1000$
test datasets using the estimated shifts from training sets. When
the results of a method give higher predictive log-likelihood value,
it indicates that the method performs better at predicting the trait
values. Figure~\ref{Fig4} shows the mean of average log likelihood values over
$1000$ test datasets with different numbers of true shifts and
coefficient sizes.

From the simulation results, when the size of coefficients are very
small or very large, methods with pBIC have a higher prediction log
likelihood value. When coefficient sizes are very small, methods with
pBIC are very strict and tend to select nearly no shifts. In these
scenarios, the signal sizes are so small that the null model has a
higher prediction likelihood compared to the true model. When
coefficient sizes are very large, all the methods can detect almost
all the true shifts, while the methods with BIC might include more
false positive shifts. Conversely, when the coefficient sizes are in
the middle of the range, methods with BIC have a better performance in
terms of prediction accuracy. PhylogeneticEM is quite conservative,
with high predictive log-likelihood when the signal sizes are
small. And in most cases, the performance of PhylogeneticEM is between
that of the pBIC and BIC methods. The difference between the ensemble
method and $\ell$1ou is smaller than the difference between BIC and
pBIC, and which method performs better vary between scenarios.

\subsection{Adjusted rand index}
An alternative approach to assess the accuracy of the chosen shifts is
to compare the induced grouping of species based on the shifts. We
used adjusted rand index (ARI, \citealp*{hubert_arabie_1985}) between
the clustering of tips of the true model and the clustering of the
estimated shifts to evaluate the model performances. The ARI is
proportional to the number of pairs in agreement between two
clusterings. The ARI is scaled and centred so that identical
clusterings give an ARI of 1 and the expected ARI of two random
clusterings is 0.

\begin{figure}

  \centering
    \includegraphics[width=\textwidth]{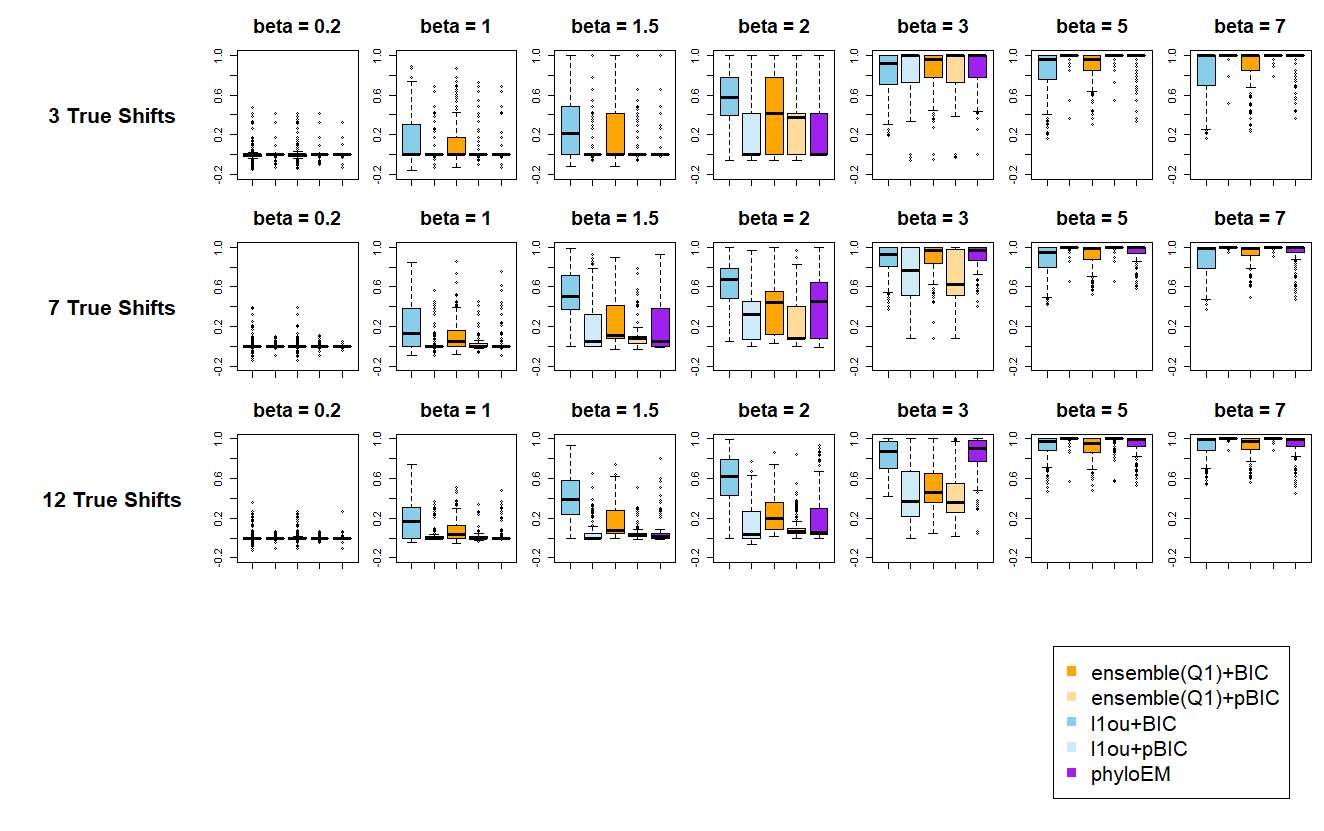}
     \caption{\label{Fig5}ARI with 3, 7, 12 true shifts}
\end{figure}

Figure~\ref{Fig5} shows the ARI comparison of the different methods
with 3, 7, and 12 shifts. ARI shows a similar result to the prediction log
likelihood. When the signal sizes are small, the methods with BIC have
a better performance. When the signal sizes are large enough, the
methods with pBIC have a higher ARI score. PhylogeneticEM has low ARI
score when the signal sizes are small and good performances when the
signal sizes are in the middle. Based on true positive versus false
positive numbers, PhylogeneticEM has poor performance with 7 and 12
shifts, but its predictive log-likelihood and ARI are comparable to
other methods. This means that the shifts estimated by PhylogeneticEM
might not be the exact true shifts, but they give similar trait
predictions and trait clustering results.

\section{Effect of shift position and tree on model performance}
\label{sec4}

The previous simulations are all based on the same phylogenetic tree
and three different shift configurations depending on the number of
shifts. In this section, we perform simulations under a wider range of
shift positions and phylogenetic trees, to assess the effect of these
factors on the performance of the various methods.

\subsection{Shifts in different positions on the tree}
\begin{figure}

  \centering
    \includegraphics[width=\textwidth]{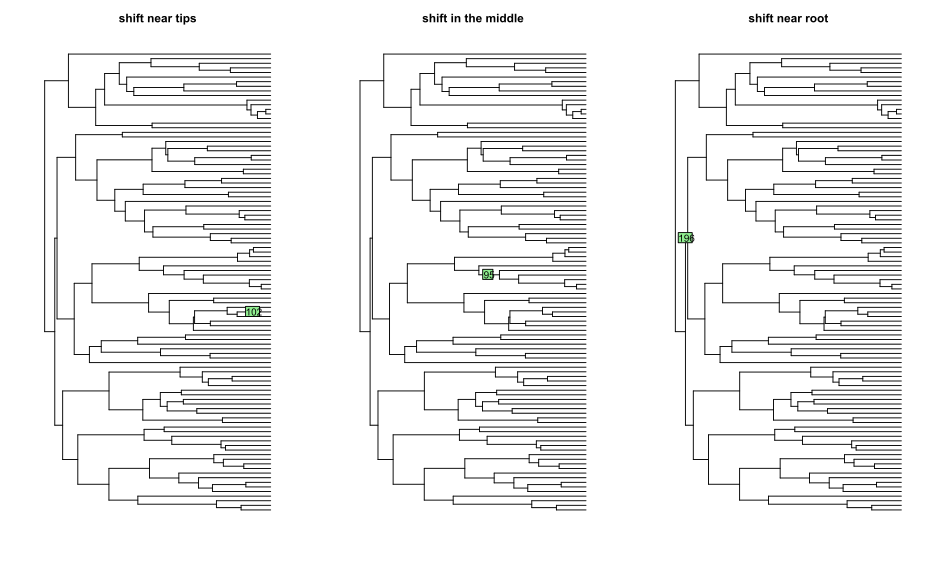}
     \caption{\label{Fig6}Shifts in different positions of tree}
\end{figure}

In this section, we study the influence of shift
position. Intuitively, different shift positions influence different
numbers of taxa and have different evolution time for the taxa so the
results might be different. Indeed the pBIC criterion is designed
specifically to account for the effect of shift position. We perform
simulations with only $1$ shift in different positions on the tree
(Figure~\ref{Fig6}). We perform simulations with $\beta=1,2,3,10$, $\alpha =1$
and $\sigma^2=2$.

\begin{figure}
  \centering
    \includegraphics[width=\textwidth]{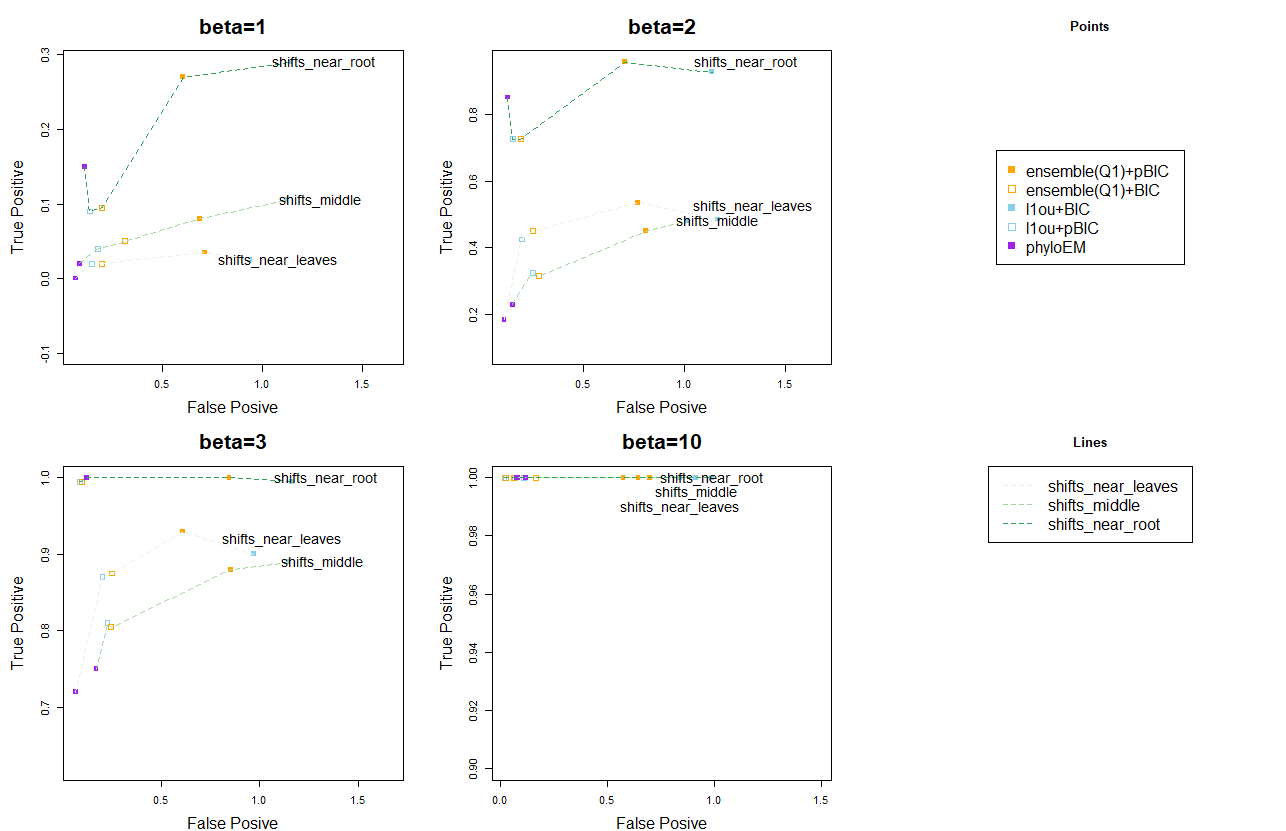}
     \caption{\label{Fig7}True positive versus false positive with different shift positions}
\end{figure}

Figure~\ref{Fig7} shows the true positive versus false positive curves
of different shift positions. When the coefficient size is very large,
all the methods can detect the true positive shift correctly,
regardless of its position. When the coefficient size is not large
enough, the shift near the root is the easiest one to detect. All the
methods have higher true positive values compared to shifts in other
positions. The result is in line with common sense --- shifts near
the root influence a large group of taxa and the evolution time after
the shift is longer, so the shift might have a larger influence on the
trait values, making it easier to detect. However, shifts near leaves
are easier to detect compared to shifts in the middle based on the
simulations for $\beta=2$ and $\beta=3$. And for $\beta=2$ and
$\beta=3$, ensemble LASSO+BIC performs better than $\ell$1ou+BIC at
detecting the shift near the root or the shift near the
leaves. PhylogeneticEM has a good performance for detecting the shift
near the root but has worse performance for detecting the shift near
the leaves and the shift in the middle.

\begin{figure}
  \centering
    \includegraphics[width=\textwidth]{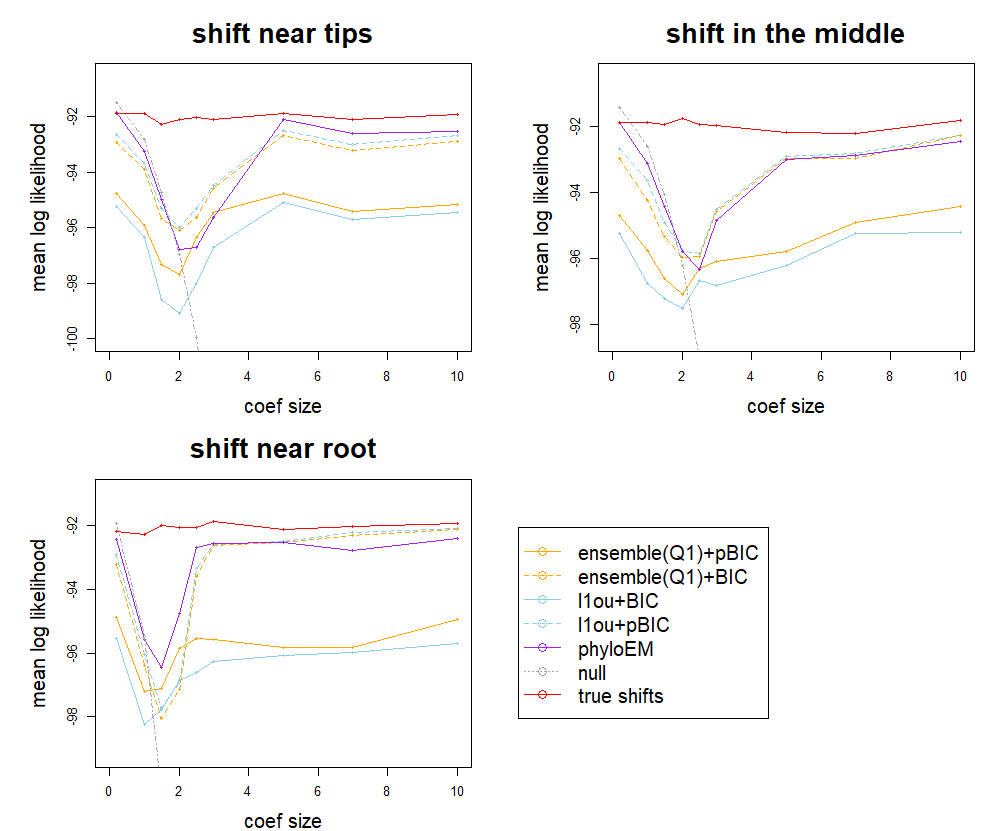}
     \caption{\label{Fig8}Log test likelihood with different shift positions}
\end{figure}

\begin{figure}
  \centering
    \includegraphics[width=\textwidth]{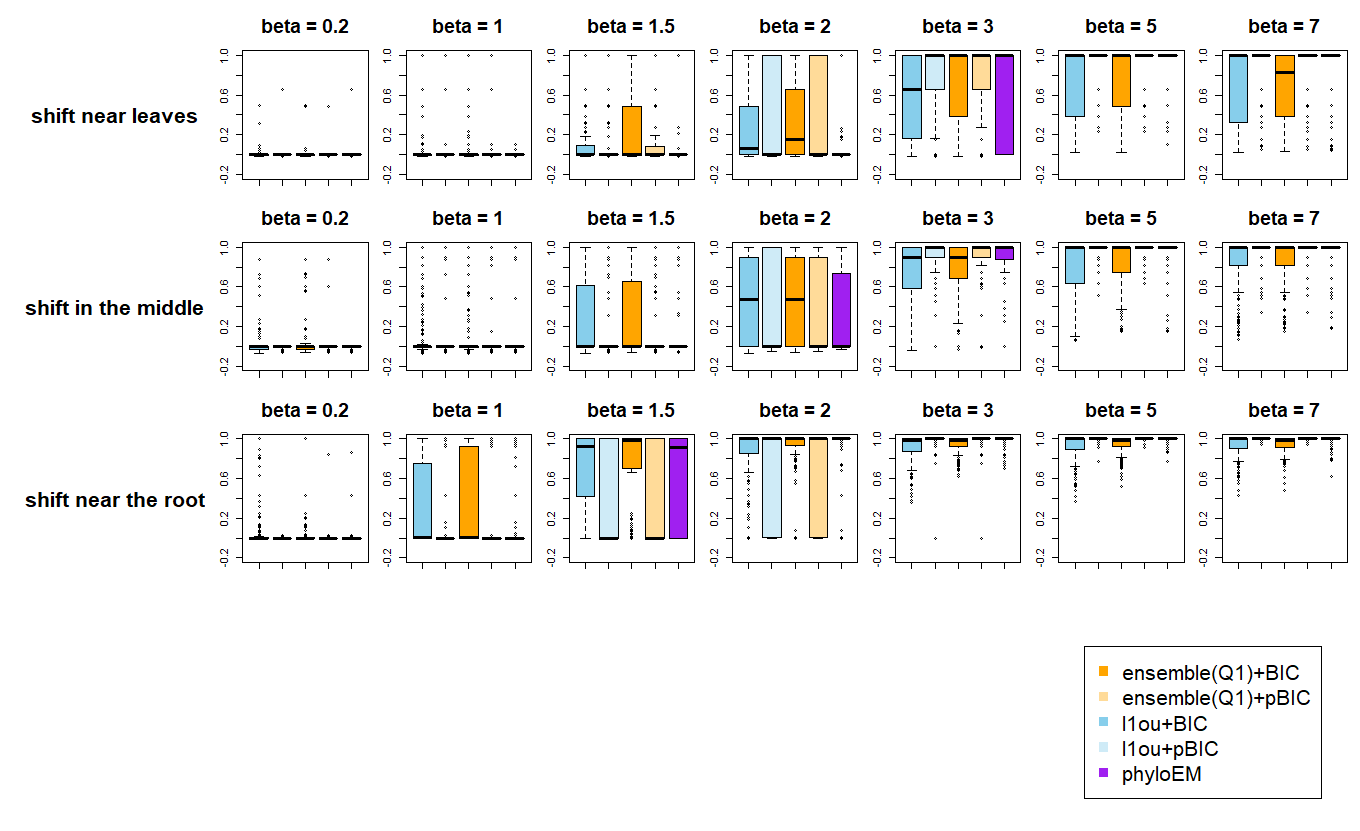}
     \caption{\label{Fig9}ARI with different shift positions}
\end{figure}

Figure~\ref{Fig8} shows the predictive log-likelihood of the methods for
different shift positions. We only include 1 shift in these
simulations, and a conservative method usually has a higher prediction
likelihood when the number of shifts is small. Thus, for the shift near
the leaves and the shift in the middle, for predictive log-likelihood,
$\ell$1ou+pBIC $>$ ensemble LASSO+pBIC $>$ ensemble LASSO+BIC $>$
$\ell$1ou+BIC. However, for the simulations with the shift near the root,
ensemble LASSO shows its strength when the coefficient size is in the
middle of the range. Figure~\ref{Fig9} shows the ARI of methods with different
shift positions. Comparing the model performances on the scenarios
with different positions of shift by ARI, all the methods perform
better with the shift near the root, then shift near the middle,
and perform worst with the shift near the leaves. This may be because
of the way ARI is defined. For a shift on a leaf branch, the ARI of a
close surrogate shift is still relatively low, whereas, for shifts near
the root or the middle of the tree, close surrogate shifts have a
significantly higher ARI.

\subsection{Different types of tree}
\begin{figure}
  \centering
    \includegraphics[width=\textwidth]{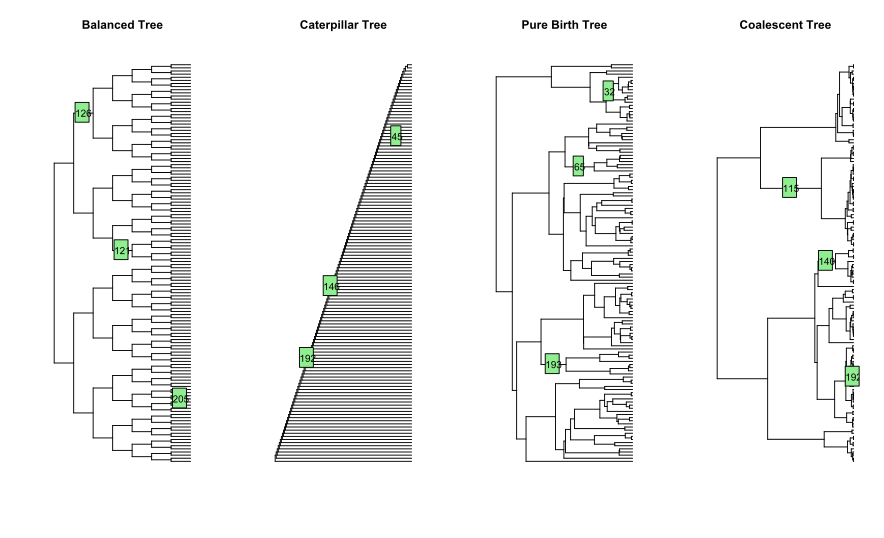}
     \caption{\label{Fig10}4 different types of tree}
\end{figure}

In this section, we compare the model performances on shift detection
tasks on different types of phylogenetic trees. We mainly consider 4
types of tree: balanced tree, caterpillar tree, pure birth tree and
coalescent tree. We generate these 4 types of trees with 128 taxa and
254 edges. Figure~\ref{Fig10} shows the generated tree and simulated shifts for
each type of tree. In this simulation, there are 3 shifts on each
tree.

\begin{figure}
  \centering
    \includegraphics[width=\textwidth]{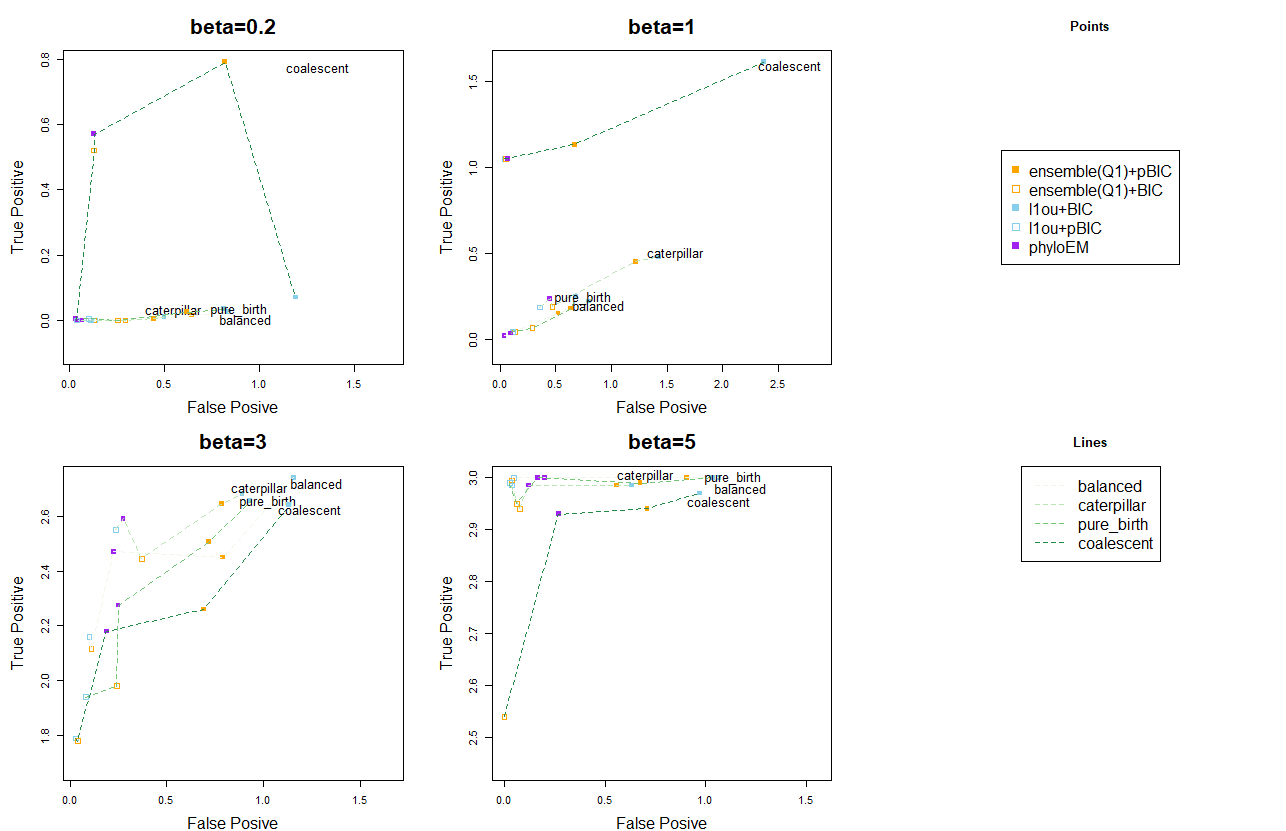}
     \caption{\label{Fig11}True positive versus false positive with different types of trees}
\end{figure}

\begin{figure}
  \centering
    \includegraphics[width=\textwidth]{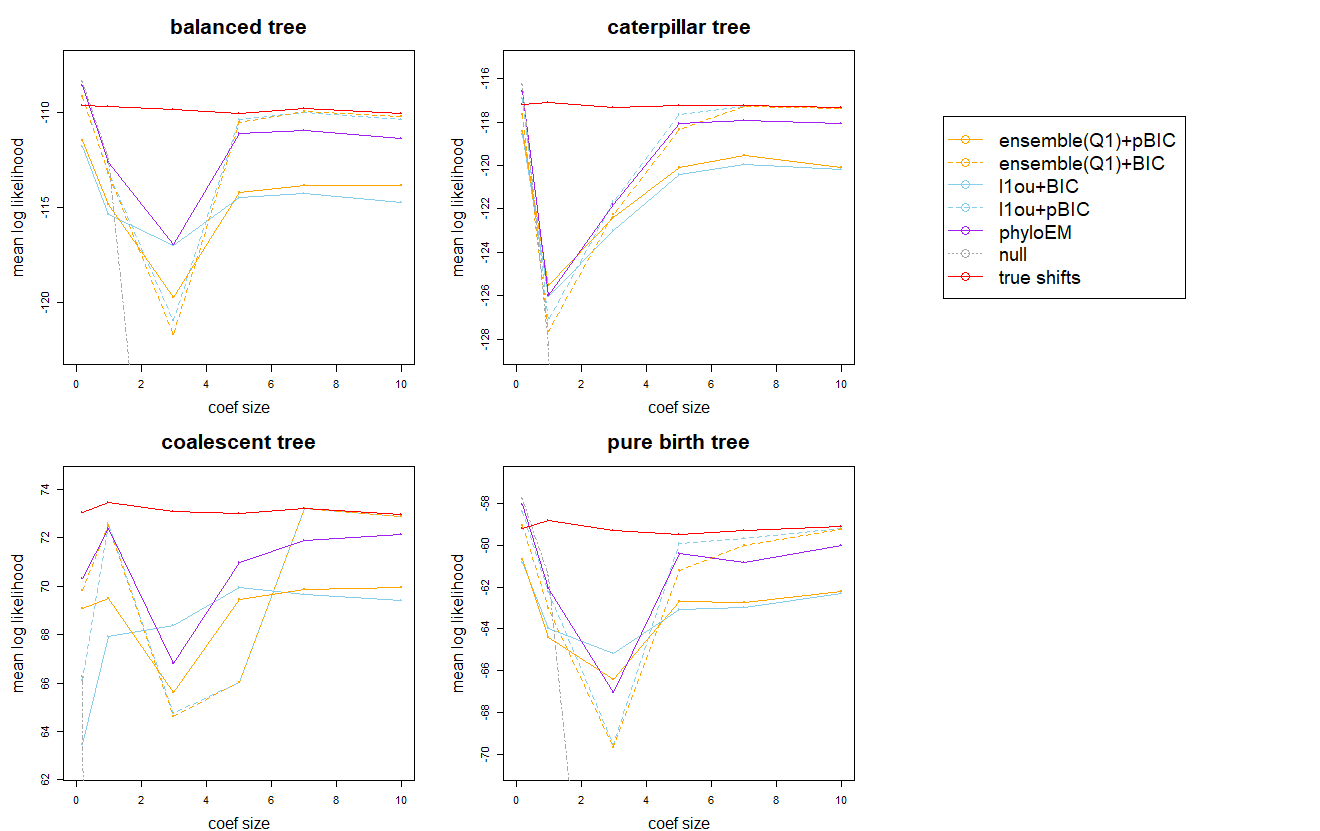}
     \caption{\label{Fig12}Average test log likelihood with different types of trees}
\end{figure}

\begin{figure}
  \centering
    \includegraphics[width=\textwidth]{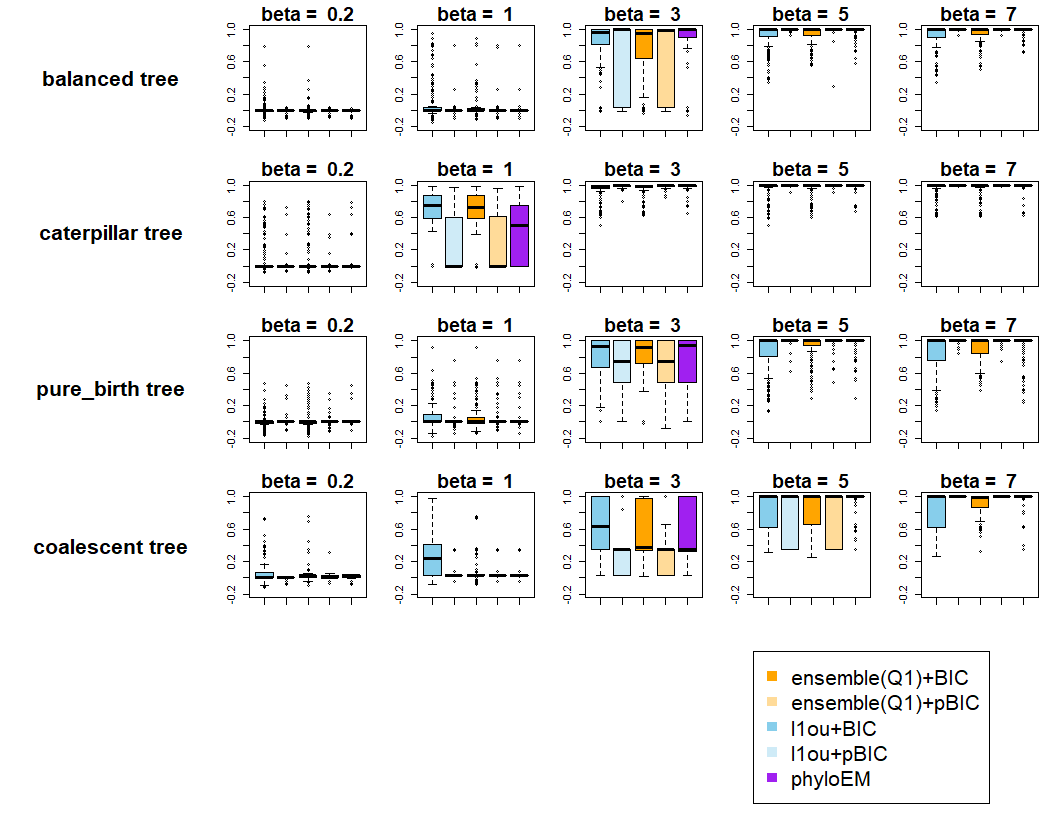}
     \caption{\label{Fig13}ARI with different types of trees}
\end{figure}

Figure~\ref{Fig11} shows the true positive versus false positive curves for
different types of trees. Interestingly, the shifts in the coalescent
tree are the easiest to detect when the coefficient size is small and
the most difficult to detect when the coefficient size is
large. Especially for ensemble LASSO, when $\beta=0.2$, ensemble
LASSO+BIC has a much higher true positive compared to other methods
and a lower false positive compared to $\ell$1ou+BIC. For other types of
tree, generally speaking when the coefficient size is in the middle of
the range, the shifts on the caterpillar tree are the easiest to
detect, and then the balanced tree, then the pure birth tree and finally
the coalescent tree. Figure~\ref{Fig12} shows the predictive log-likelihood. It
is obvious that ensemble methods perform very well on the coalescent
tree with small signal size and poorly with large signal
size. Figure~\ref{Fig13} shows the ARI with different types of trees. From the
ARI results, the order of difficulties of shift detection on
different types of trees is coalescent tree $>$ pure birth tree $>$
balanced tree $>$ caterpillar tree.


\section{Robustness to violations of model assumptions}
\label{sec5}

The simulations above assume that the model assumptions in
Section~\ref{sec2} are met.  However, in reality, the real data often
violate the model assumptions. For example, the trait values might
contain measurement errors. Measurement errors enlarge the variance of
the data and make accurate detection more difficult. Another example
is that the phylogenetic tree is constructed based on sequence data,
thus can also have errors. These kinds of problems are hard to avoid in
real data analysis. It is, therefore, important to consider how the
methods perform when the model assumptions are not satisfied. In this
section, we discuss several violations of model assumptions. We also
study the effects of misestimation of parameters on model selection
performance. Because the $\ell$1ou and ensemble methods use a very
crude approach to estimate $\alpha$, it is important to examine the
question of how misestimation of $\alpha$ impacts results.
PhylogeneticEM does not suffer from this problem because it does not
have a preconditioning step.

\subsection{Measurement errors}
It is common for the measurement of traits to be subject to errors. These
errors may impact the shift detection methods, which assume that the
trait values are measured perfectly. In this section, we simulate
additive Gaussian measurement error $N(0,\sigma_e^2)$ for the trait value of each species. When $\sigma_e^2$ is larger, the size of the
measurement error is larger.

\begin{figure}
  \centering
    \includegraphics[width=\textwidth]{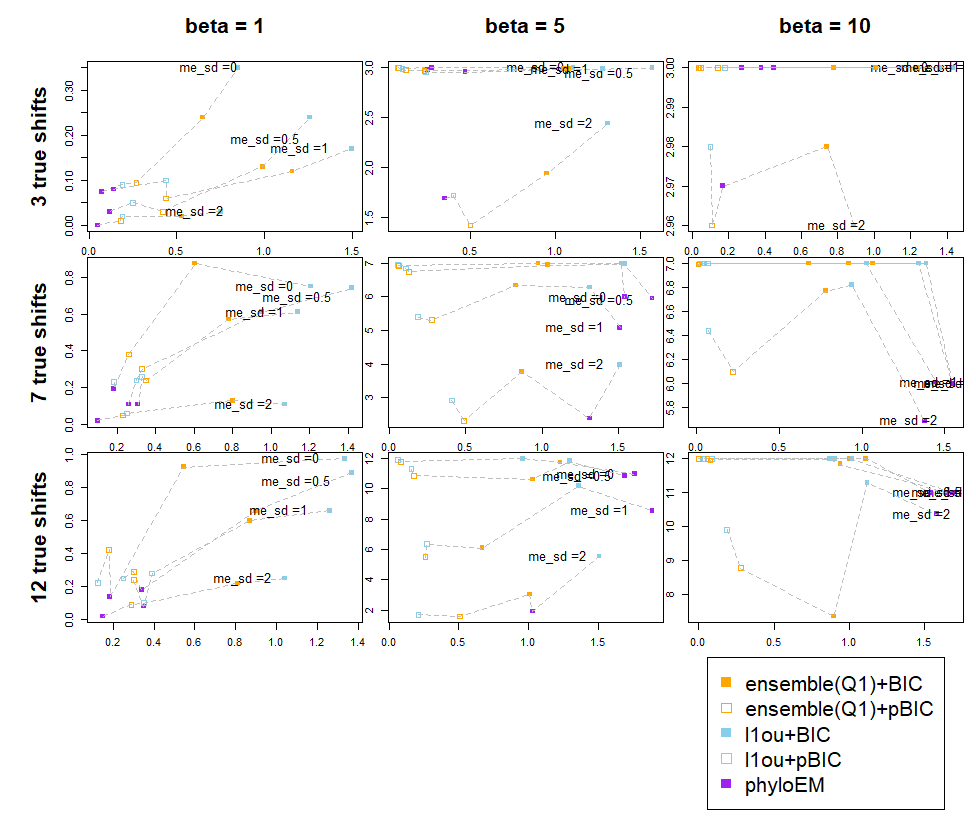}
     \caption{\label{Fig14}True positive versus false positive with changing measurment error.}
\end{figure}

\begin{figure}
  \centering
    \includegraphics[width=\textwidth]{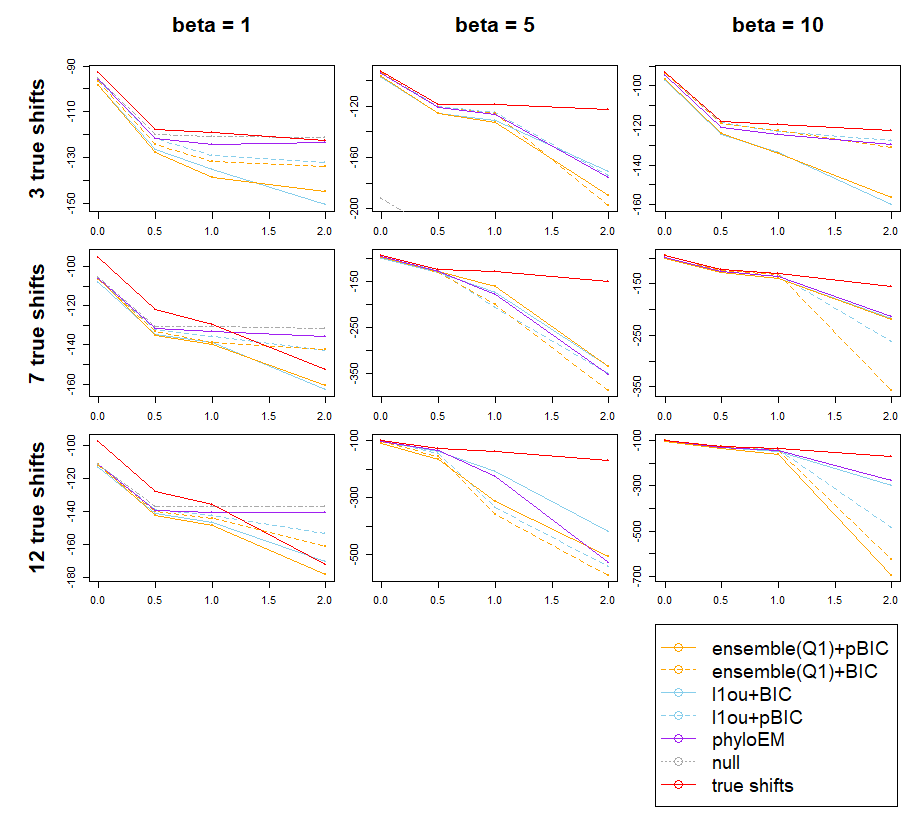}
     \caption{\label{Fig15}Average test log likelihood with parameters estimated from training data with measurement error}
\end{figure}

\begin{figure}
  \centering
    \includegraphics[width=\textwidth]{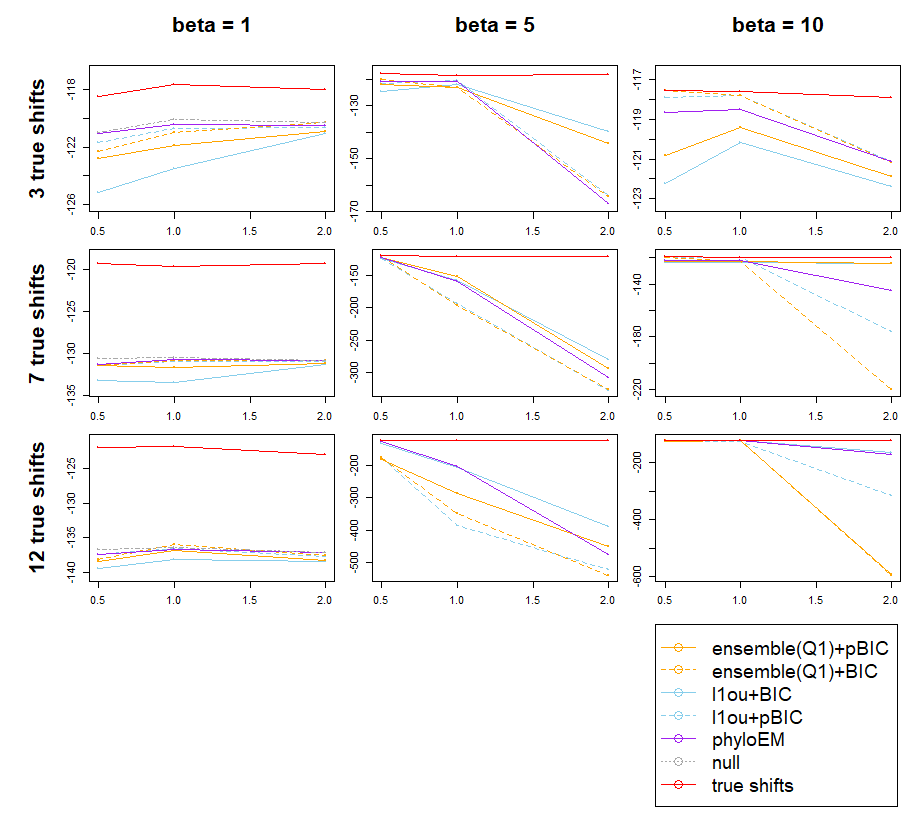}
     \caption{\label{Fig16}Average test log likelihood with parameters estimated from training data without measurement error}
\end{figure}

\begin{figure}
  \centering
    \includegraphics[width=\textwidth]{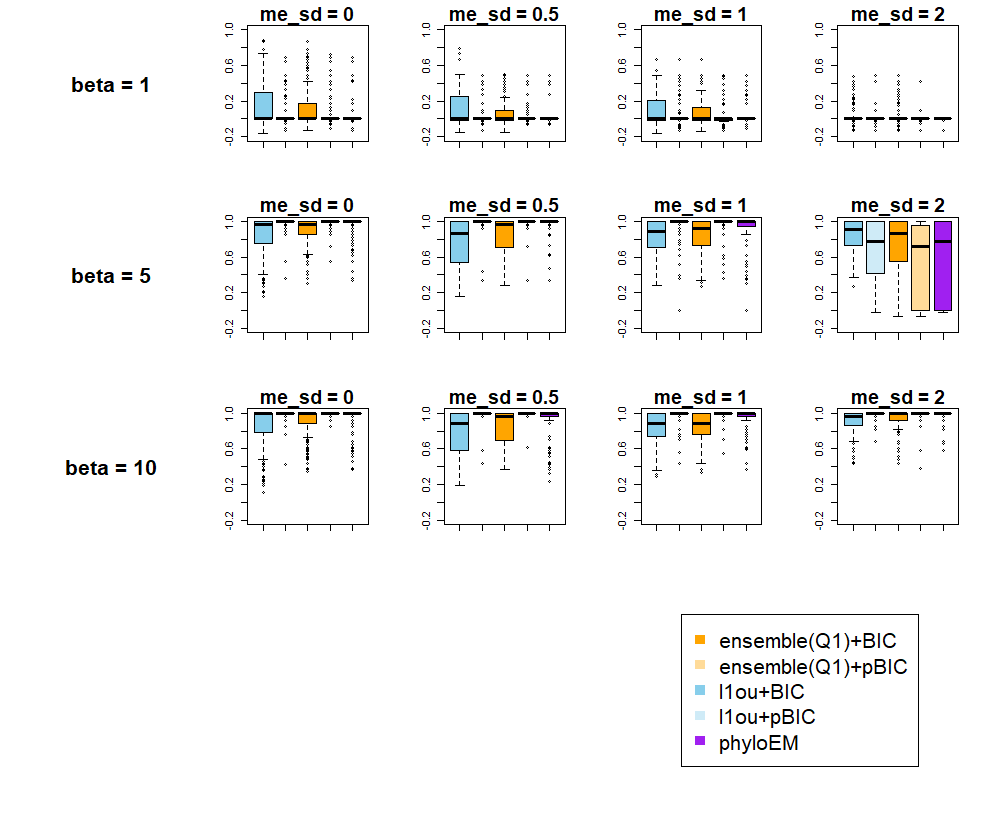}
     \caption{\label{Fig17}Distribution of ARI between selected shift configuration and true shift configuration with changing measurment error.}
\end{figure}

Figure~\ref{Fig14} shows the true positive versus false positive with
changing measurement errors. Figure~\ref{Fig15} shows the predictive
log-likelihood value with changing measurement errors in training
data. Because of the measurement errors in the training data, the true
shift model has a lower log-likelihood than the null model.  If
we use the training data with measurement error to estimate the
parameters of selected shifts, the parameters will be far from the
real values and thus introduce errors while calculating the predictive
log-likelihood. Since our purpose is to identify the true shifts,
predictive log-likelihood using training data with measurement error
to estimate the parameters is not an ideal way to assess the selected
shifts. Therefore, we also compare the prediction log-likelihood of the
shift detection methods using the training data with measurement error
to select shifts, but with parameters estimated from training data
without measurement error.
Figure~\ref{Fig16} shows the results of log likelihood with parameters
estimated from training data without measurement error.
Figure~\ref{Fig17} shows the ARI with measurement error. The plots show
that the measurement error will influence the accuracy of shift
detection. The loss of accuracy increases with the number of
shifts. The performance of all the methods worsens with measurement
error. When the signal is not so strong ($\beta=1$), even a relatively
small measurement error severely impacts performance. Interestingly,
in terms of ARI, when $\beta=10$, the performance of the methods using
BIC improves with increased measurement error. Looking at
the true and false positive rates, we see that the methods
consistently select the true variables, but with increased measurement
error, the false positive rate is reduced. However, the predictive
log-likelihood does decrease. This may be because even though failing
to select a true variable is very rare in this case, it has a huge
impact on the log-likelihood, because of the strong signal
strength. Thus the effect on predictive log-likelihood may be driven
by the small number of cases where a true variable is missed.


\subsection{Brownian-Motion model}
Another assumption that may be violated is that the evolution on each
branch follows an OU process. In this section, we use BM to simulate the trait values. This is a very simple simulation of this
violation since BM is a special case of the OU process
with $\alpha=0$. Other models should be considered in future studies.\

\begin{figure}
  \centering
    \includegraphics[width=\textwidth]{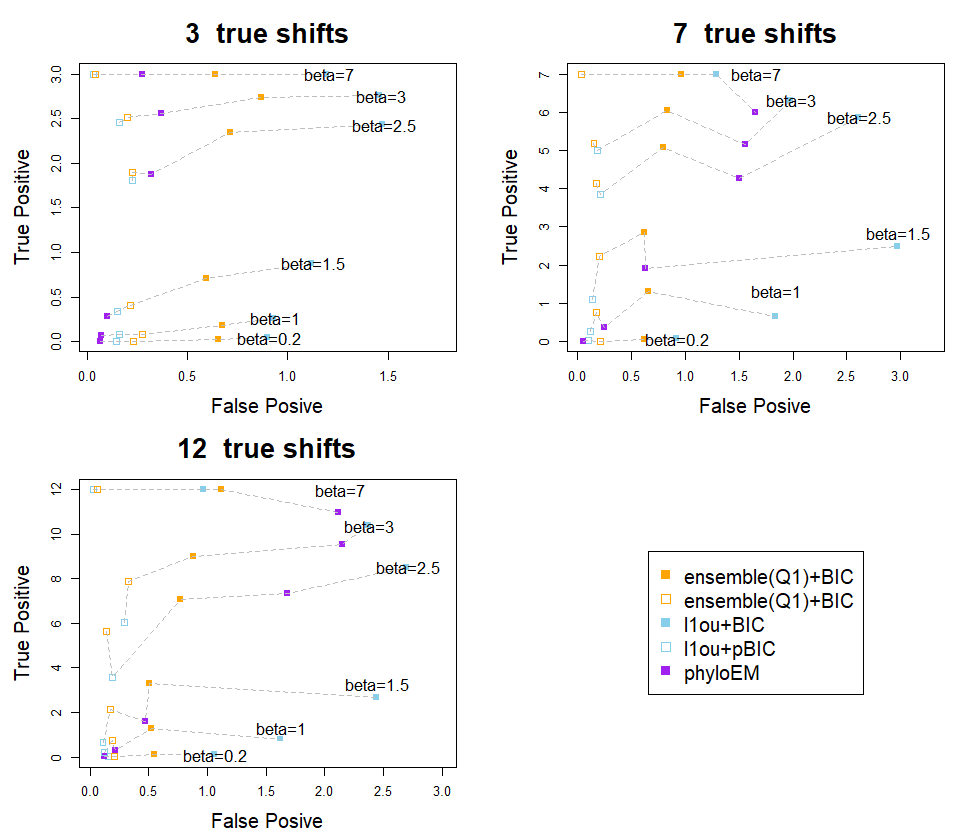}
     \caption{\label{Fig18}True positive versus false positive with BM model.}
\end{figure}

\begin{figure}

  \centering
    \includegraphics[width=\textwidth]{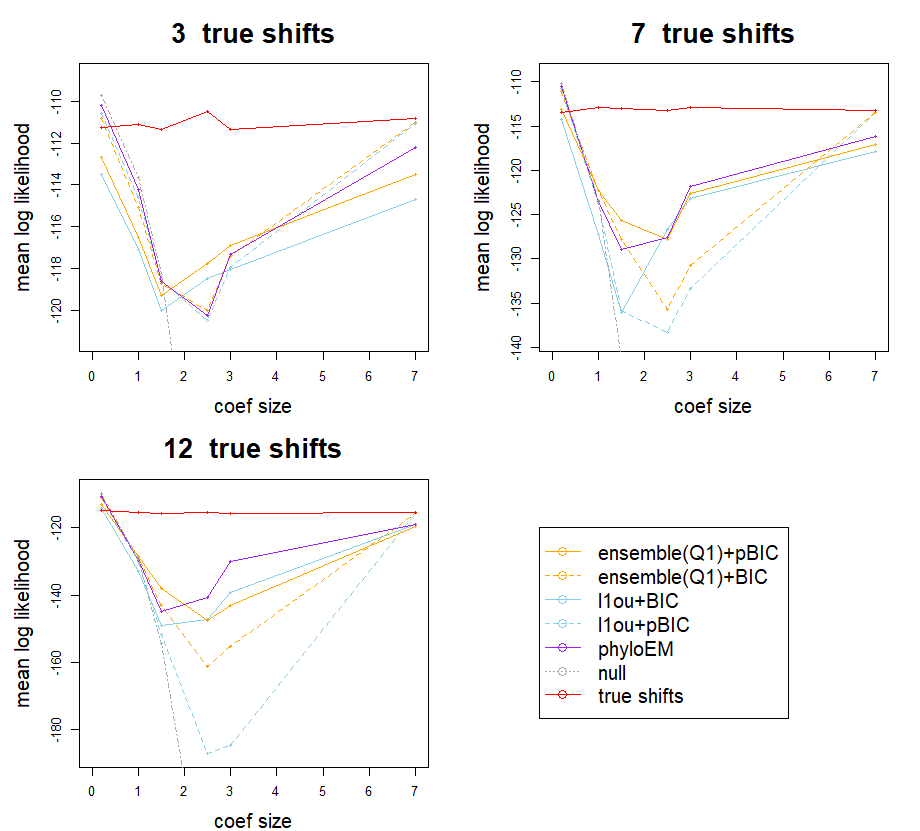}
     \caption{\label{Fig19}Average test log likelihood with BM model.}
\end{figure}

\begin{figure}
  \centering
    \includegraphics[width=\textwidth]{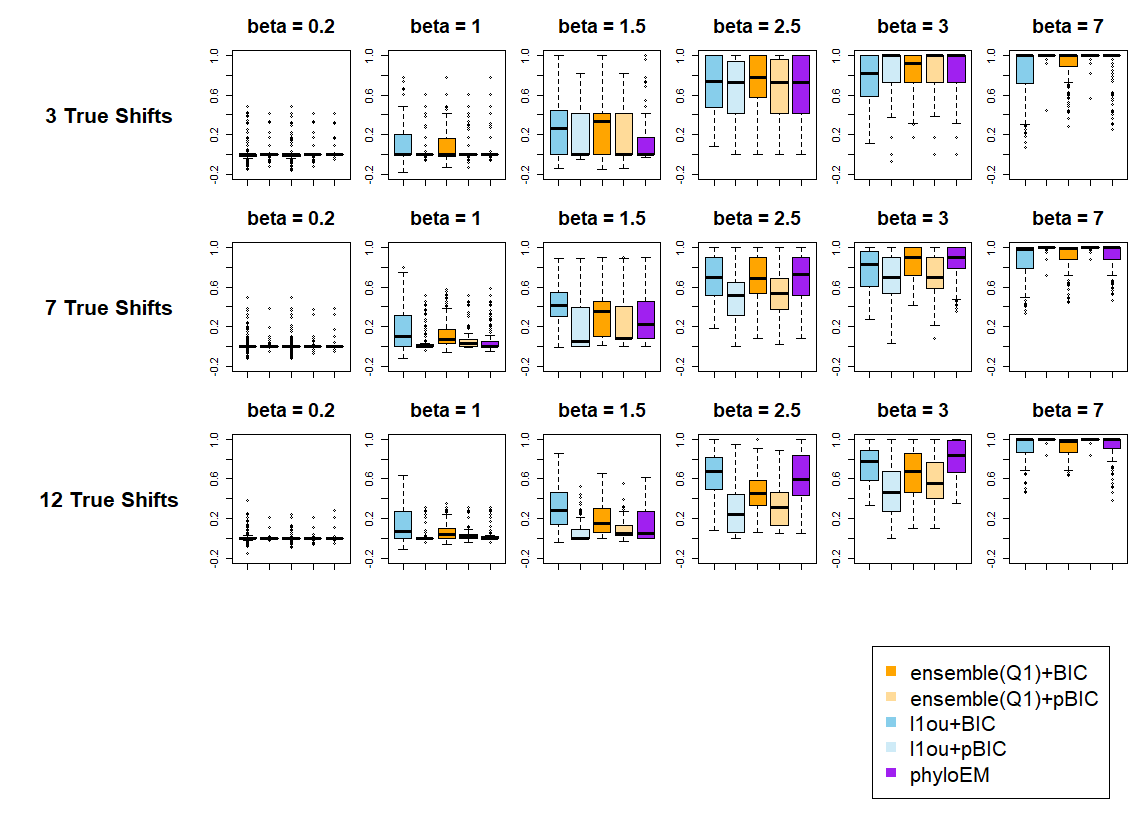}
     \caption{\label{Fig20}ARI with BM model. 3 true shifts simulations are shown in this figure.}
\end{figure}

Based on the results of Figures~\ref{Fig18},~\ref{Fig19},
and~\ref{Fig20}, under the BM model, the results are similar compared
to the results under the OU model. There is no significant failure of
any method when applied to data generated from the BM model. The
strength of ensemble LASSO methods when there are 7 or 12 shifts
becomes more obvious. Ensemble LASSO+BIC has a lower false positive
rate compared to $\ell$1ou+BIC and similar or even higher true
positive rate. Ensemble LASSO+pBIC has a higher true positive rate
compared to $\ell$1ou+pBIC and similar or even lower false positive
rate.


\subsection{Incorrect tree}\label{MisspecifiedTree}
\begin{figure}

  \centering
    \includegraphics[width=\textwidth]{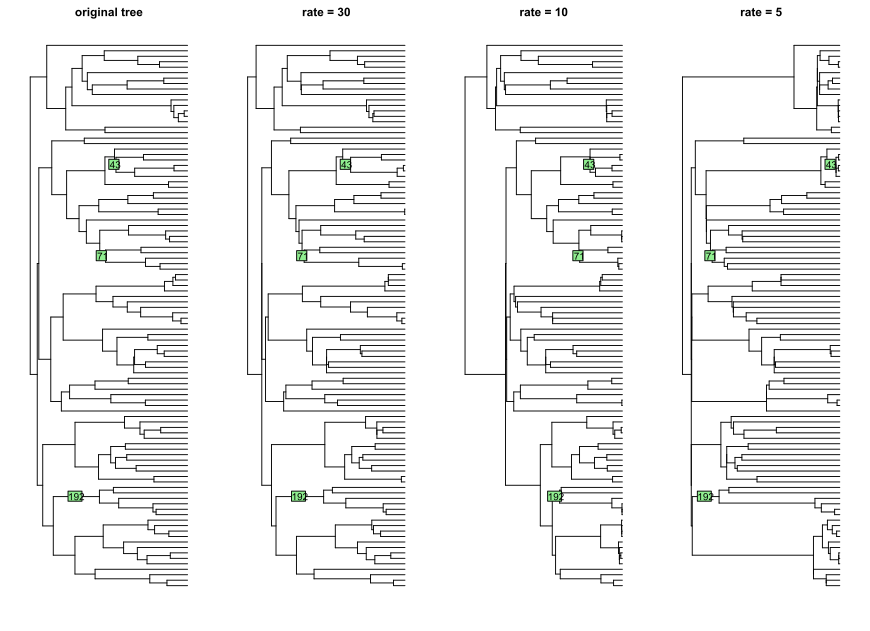}
     \caption{\label{Fig21}Regenerated tree with different beta}
\end{figure}

Another possible violation is that we use a wrong tree to do the shift
detection. There are two possible errors in the analyzed tree: wrong
topology or wrong branch lengths. In the case of an incorrect
topology, it is not always clear that there is a meaningful way to
define ``true shifts'' in a false tree. Therefore, we will focus on
the case with incorrectly specified branch lengths. We simulate data
from a tree with different branches and apply the methods to the
originally provided tree. We use a gamma distribution
($Gamma(\beta*\mathrm{original\ branch\ length},\beta)$) to randomly
generate each internal branch length. Thus, the mean of each branch
length is the original branch length, and the variance is larger for
smaller $\beta$. In order to generate ultrametric trees and keep the
total tree depth the same as the original tree, the external branches
are generated by the original tree depth minus the tree depth of the
starting node of each external branch. The internal branches are
generated with a depth first order. If the depth of one internal
branch is larger than the given tree depth, this branch length is
resampled. We generate 3 trees with $\beta = 30,10,5$ as shown in
Figure~\ref{Fig21}.

\begin{figure}

  \centering
    \includegraphics[width=\textwidth]{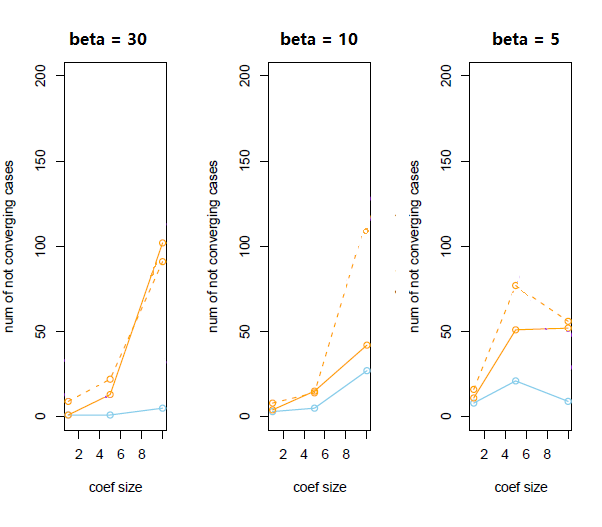}
     \caption{\label{Fig22}Number of non-converging cases while analyzing on a misspecified tree }
\end{figure}

Firstly, we find that this violation of assumptions can cause
convergence problems while iterating the methods to estimate $\alpha$
in $\ell$1ou or the ensemble method. This is a problem we
observe for real data in Section~\ref{real_data}. Figure~\ref{Fig22}
shows the number of cases where the estimated alpha does not converge
in 10 iterations out of 200 simulations. The convergence problems seem
to particularly affect the ensemble method but are also present for
$\ell$1ou. For the non-converging simulations, we present results for
the best $\alpha$ value attempted (in terms of our model selection
criterion).

\begin{figure}

  \centering
    \includegraphics[width=\textwidth]{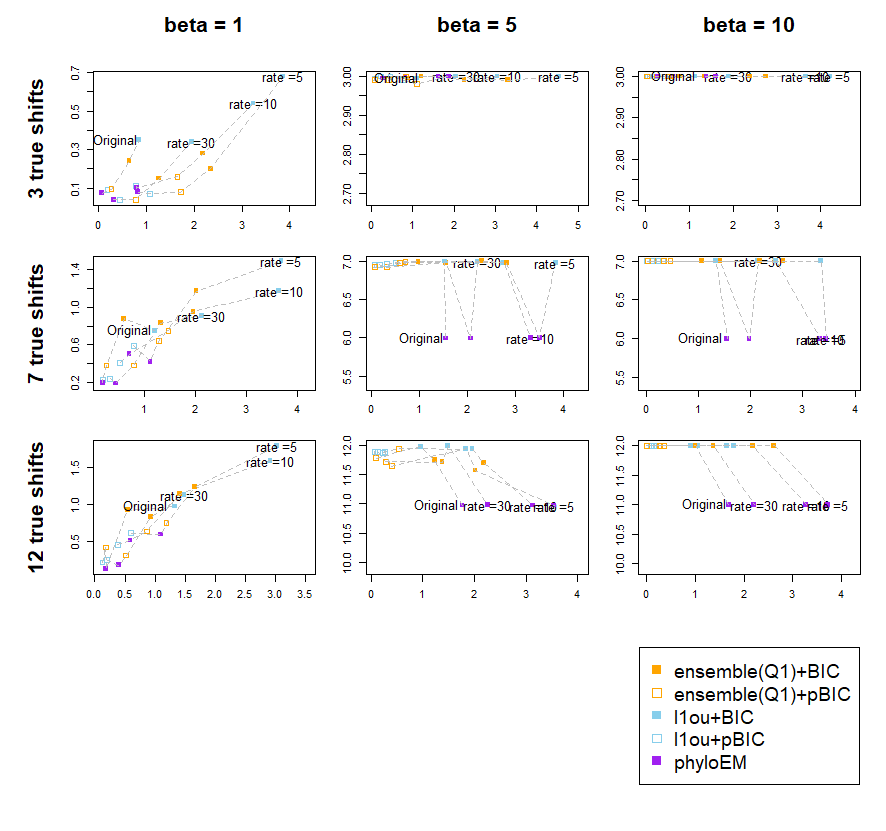}
     \caption{\label{Fig23}True positive versus false positive with applying methods on misspecified trees}
\end{figure}

\begin{figure}
  \centering
    \includegraphics[width=\textwidth]{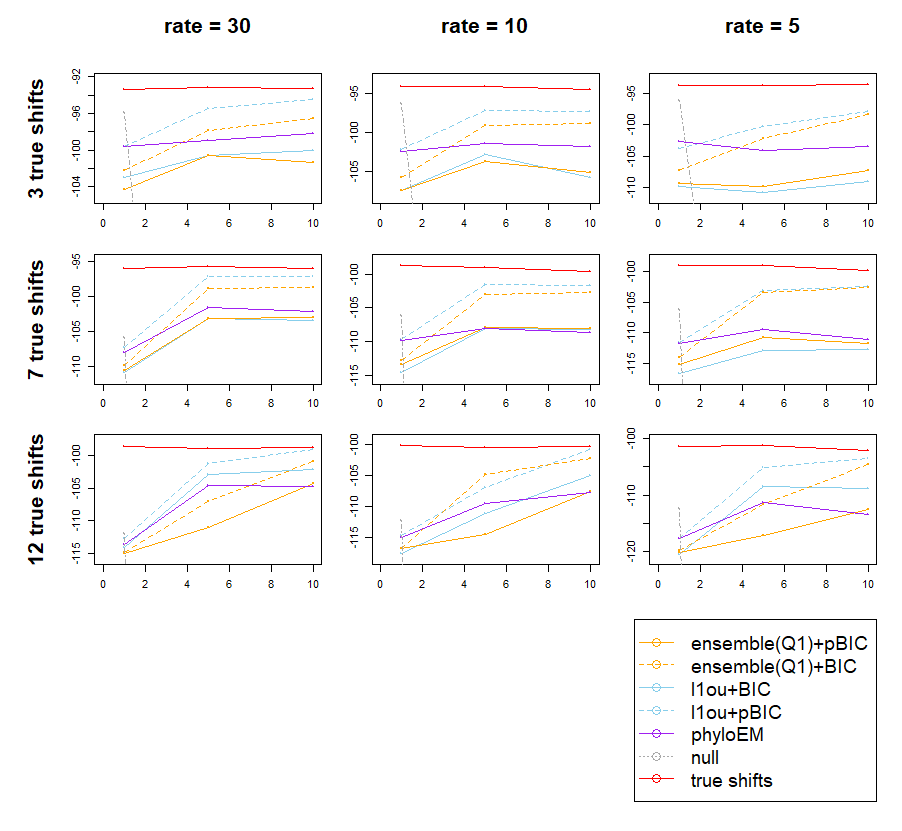}
     \caption{\label{Fig24}Average test log likelihood with applying methods on misspecified trees}
\end{figure}

\begin{figure}
  \centering
    \includegraphics[width=\textwidth]{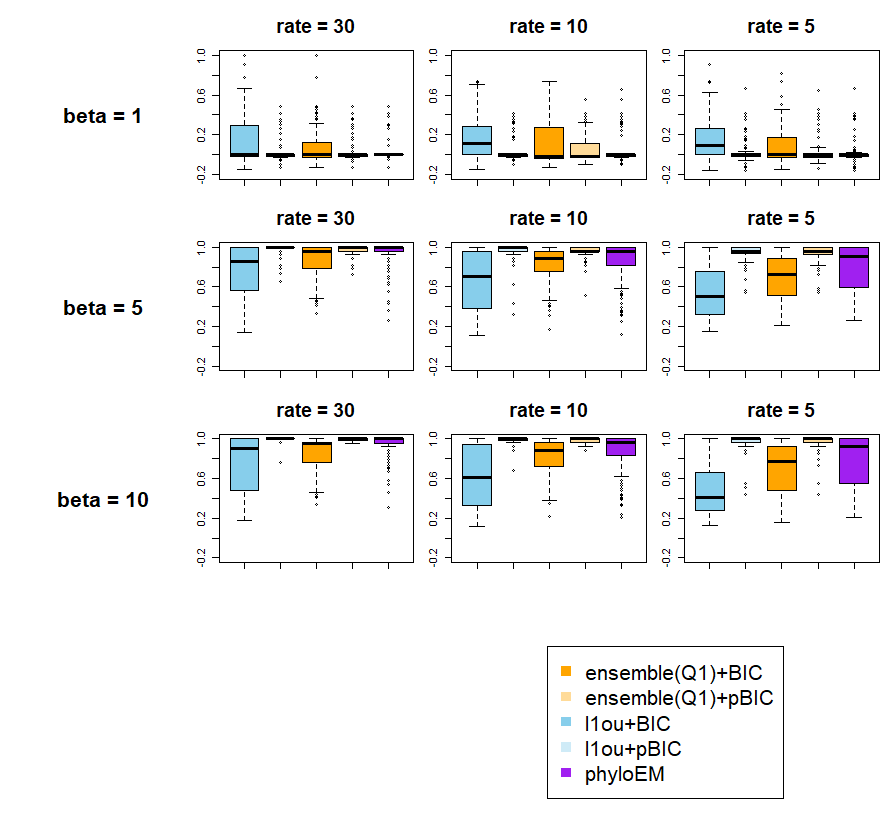}
     \caption{\label{Fig25}ARI with  methods applied on misspecified trees}
\end{figure}

Figure~\ref{Fig23} shows the average number of true and false positives for
each misspecified tree. Figure~\ref{Fig24} shows the prediction
log-likelihood. Figure~\ref{Fig25} shows the ARI between the estimated and true
shift configurations.  We see that generally speaking, the difference
between the real tree and the analyzed tree will worsen the performance of
the methods by either lowering the true positive rate or increasing
the false positive rate. However, the methods are relatively robust to
this misspecification. PhyloEM appears to be most influenced by the
misspecification, becoming less conservative when the tree is
misspecified. This increases both the true positive rate and the false
positive rate. The predictive log-likelihood on test data is fairly
robust to the tree misspecification. However, the ARI decreases,
particularly for the cases with strong signals, as the tree becomes
more misspecified.

In some exceptional cases, the methods perform better when
the tree is misspecified. For example, for simulations with 7 or 12
true shifts, nearly all the methods got higher true positive rates
when the tree was most misspecified (rate parameter = 5) than for the
original tree.

\subsection{Misestimation of $\alpha$}
Recall from Section~\ref{sec2} that both $\ell$1ou and ensemble use a very rough
method to estimate $\alpha$. This could lead to the estimated $\alpha$
values being very bad. In this section, we investigate the extent to
which misestimation of $\alpha$ can impact the variable selection
results. In this simulation, we use $\alpha = 1$ to generate data and
use different $\hat\alpha$ $(10^{-4},10^{-3},...,10^2)$ in the methods
and compare the model performances.

Figure~\ref{Fig24} and Figure~\ref{Fig25} show the prediction log
likelihood and ARI with different estimated $\hat{\alpha}$. From the
plots, the performance of PhylogeneticEM is influenced most by
changes in the estimation of $\alpha$. Especially when the estimated
$\alpha$ is too large, the method performs poorly. Ensemble methods
and $\ell$1ou are more robust to misestimation of $\alpha$. Since PhlyoEM
uses maximum likelihood to estimate $\alpha$, which is expected to
produce more accurate estimates, robustness to misestimation is less
important than for $\ell$1ou and the ensemble method, which use a very
rough method to estimate $\alpha$.

\begin{figure}
  \centering
    \includegraphics[width=\textwidth]{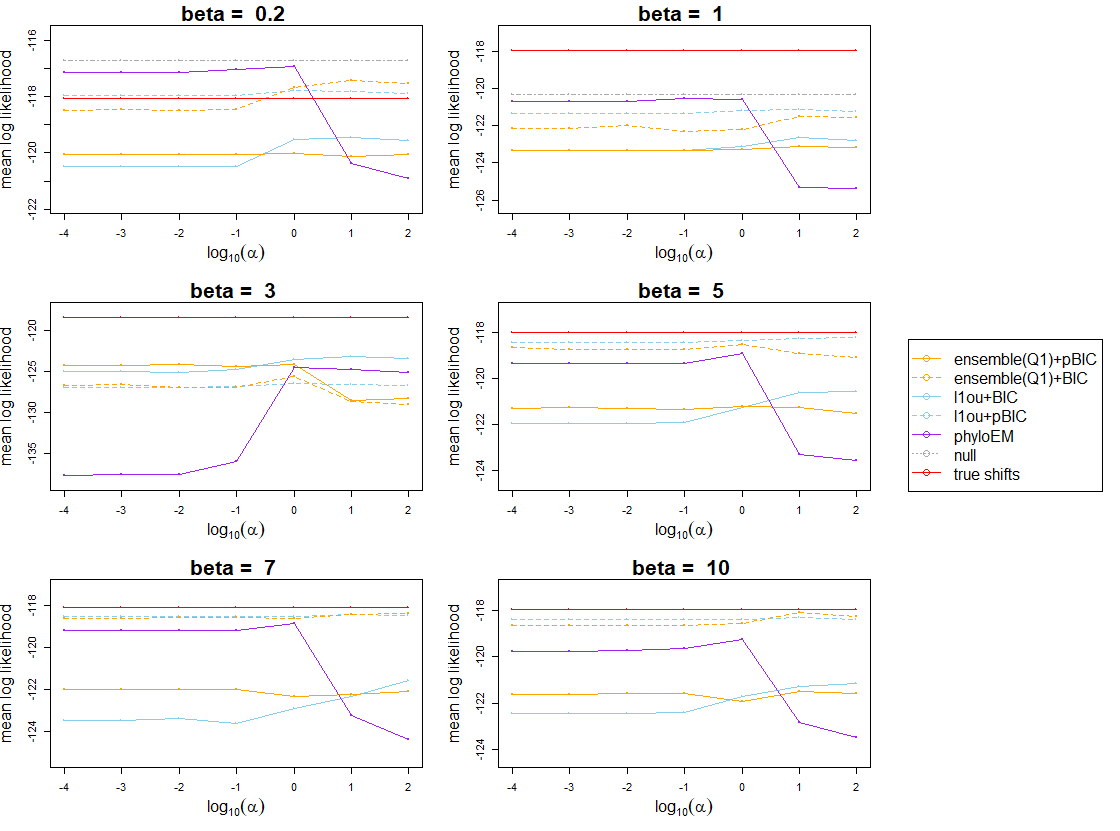}
     \caption{\label{Fig26}Average test log likelihood with changing estimated alpha}
\end{figure}
\begin{figure}
  \centering
    \includegraphics[width=\textwidth]{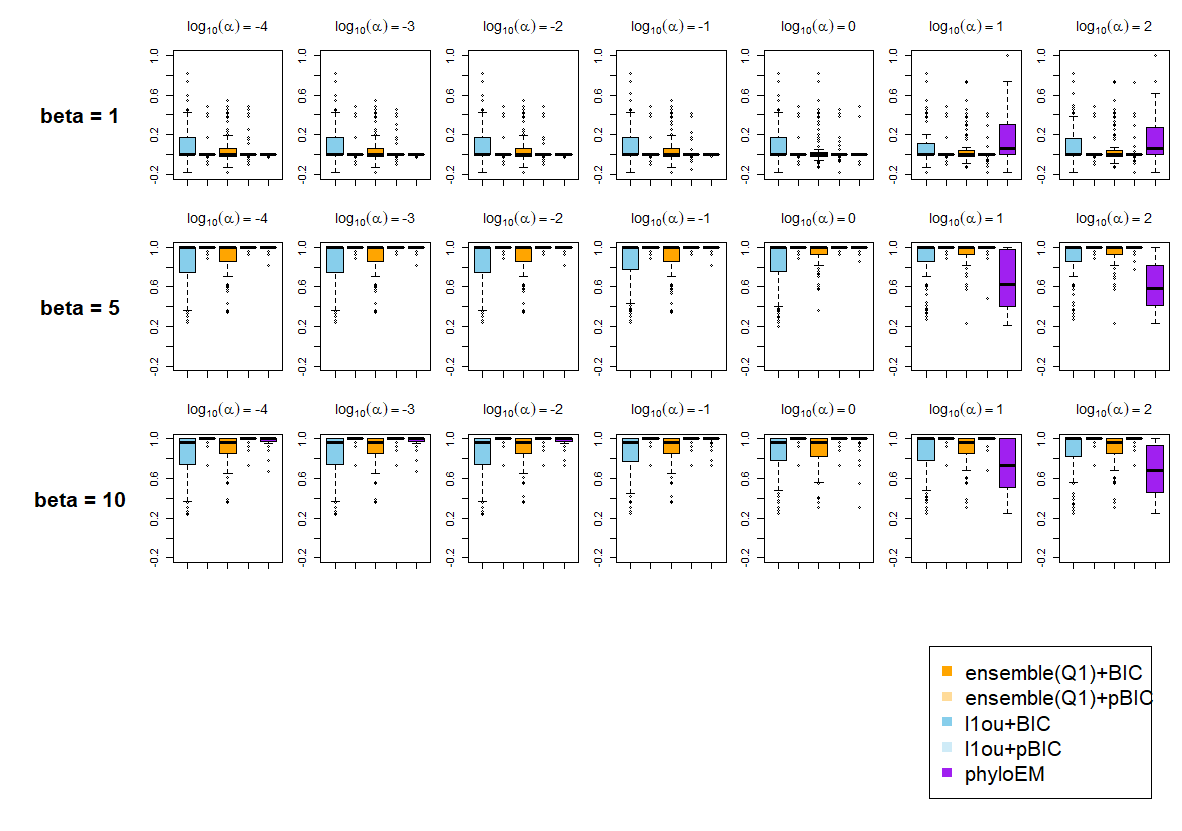}
     \caption{\label{Fig27}ARI with changing estimated alpha}
\end{figure}

\section{Case study: Anolis Lizard data}\label{real_data}
The tree and trait data of Anolis lizards are provided by
\citet{mahler_ingram_revell_losos_2013}.  They applied PCA to 11
traits including body size, limb, tail length, and so on. They found
that the first four principal components explained 93\% of the
variation. The data are available in the \texttt{R} package
$\ell$1ou. We compare the results of the different variable selection
methods for the Anolis Lizard data.

\subsection{Results}

\begin{figure}

  \centering
    \includegraphics[width=\textwidth]{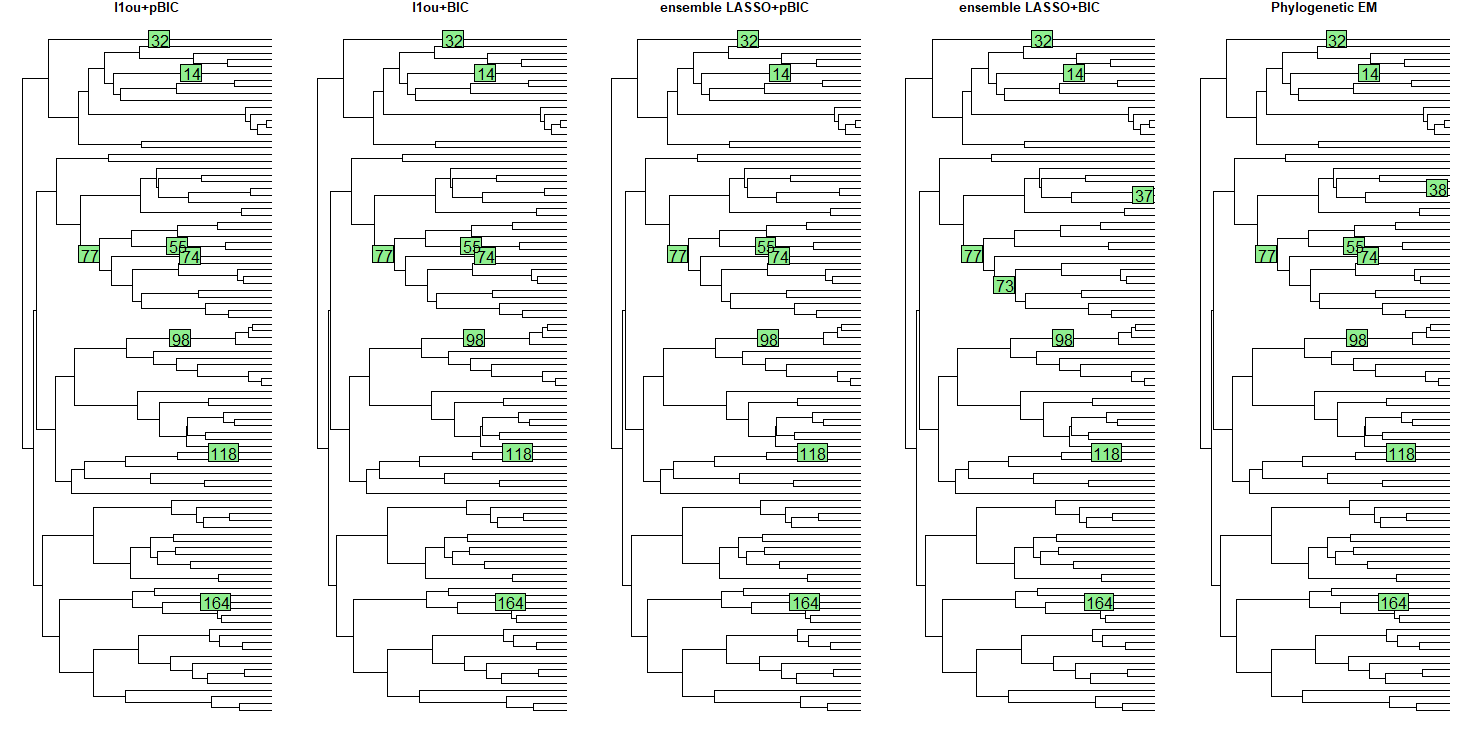}
     \caption{\label{Fig28}Shifts detected by different models for lizard data}
\end{figure}

Figure~\ref{Fig27} shows the shifts detected by applying
$\ell$1ou+pBIC, $\ell$1ou+BIC, ensemble LASSO+pBIC, ensemble
LASSO+BIC, and PhylogeneticEM on the first principal component of the
lizard traits. All the methods have similar results. $\ell$1ou+pBIC,
$\ell$1ou+BIC and ensemble LASSO+pBIC detected the same
shifts. Ensemble LASSO+BIC has slightly different results, only
differing for a small number of species. Both ensemble LASSO+BIC and
PhylogeneticEM detect one more shift near the leaf node that the other
three methods do not select. The estimated shift sizes range from
$-5.50$ to $-1.91$. By applying several methods to the same dataset,
we might say that the shifts which are selected by all the five
methods are highly likely to be true. While the shifts which are
selected by only one or two methods might need further investigation
and more biological evidence.

There are also convergence problems for the estimation of $\alpha$ for
this real dataset.  From Section~\ref{MisspecifiedTree}, this might
indicate that the tree is misspecified, or some other violation of the
model assumptions.

\subsection{Recommendations}
We suggest applying multiple methods to each data set and comparing
the results. Based on the simulation results, different methods have
different strengths and we cannot say that any method outperforms the
others in every situation. For example, the methods with pBIC are the
most conservative: they perform best with large signal sizes. They can
detect the true shifts without introducing false positive
shifts. However, they cannot give reasonable results when the signal
sizes are small. The ensemble method with BIC can better capture the
shifts near the leaves. PhylogeneticEM is even more conservative with
small signal sizes and falls between methods with pBIC and with BIC,
with large signal sizes. It is hard to tell which method and criterion
is the most suitable to use in a specific task. By comparing the
results of different methods, we can get the confidence level of the
selected shifts. For example for the Anolis lizard data, the shifts
which are selected by all the five methods are highly likely to be
true. \citet{khabbazian_kriebel_rohe_ane_2016} use bootstrap support
to evaluate how likely the selected shifts are true. However, the
bootstrap support can be influenced by biases in a particular
method. By combing the results of several different methods, we can
assess the confidence of particular shifts in a way that is unlikely
to be influenced by the bias of any particular method.

\section{Conclusion}
\label{sec7}
In this article, we compared the performances of several shift
detection methods --- $\ell$1ou, PhylogeneticEM, ensemble method --- for
trait evolution models. To understand the strength, weaknesses, and
restrictions of different methods, we compared the performances over a
large range of scenarios. We used three different measurements to
compare the results, true positive versus false positive curve,
predictive log-likelihood and adjusted rand index. 

From the simulation results, when the coefficients are very small,
PhylogeneticEM, $\ell$1ou+pBIC and ensemble+pBIC are very strict and
tend to select nearly no shifts. In these scenarios, ensemble+BIC and
$\ell$1ou+BIC perform better at detecting the small magnitude
shifts. However when the coefficients are large, nearly all the
methods can detect the true shifts, but $\ell$1ou+BIC and ensemble+BIC
include more false positive shifts. The performances of methods are
highly dependent on the criterion. A better criterion might help the
methods to give very good results with varying signal sizes. Further
research about appropriate model selection criteria for shift detection
might be an interesting topic for future studies.

Furthermore, we compared the model performances on different shift
positions in trees and different types of trees. From the results, the
shifts near the leaves are the most difficult to detect and the shifts
near the root are the easiest to detect. The shifts on the coalescent
tree are the easiest to detect when the coefficient is small and the
most difficult to detect when the coefficient is large.

We also conducted simulations in several scenarios where the model
assumptions do not hold. We studied training data with measurement
error; misspecified phylogenetic trees; a non-OU model; and
misestimation of the parameter $\alpha$. From the simulation results,
measurement error and a misspecified phylogenetic tree cause shift detection
more difficult and all the methods perform worse in these
cases. $\ell$1ou and the ensemble method are robust to misestimation
of $\alpha$. The simulations under a BM model in this article do not
cause much difficulty. However, further research into different model
misspecifications is still necessary.

In the Anolis Lizard real data case study, all five methods estimate
similar shift configurations. There are only a few differences in the
shifts near the leaves. The shifts which are selected by all the five
methods are highly likely to be true, while the shifts which are
selected by only one or two methods might need further
investigation. Interestingly, we observe convergence problems for the
estimation of $\alpha$ for this real data with $\ell$1ou and the ensemble
method. In our simulations, estimation on training data with
a misspecified tree often led to this convergence problem, which
suggests that the Anolis Lizard tree may be misestimated.

\section{Data accessibility}
The tree and trait data of Anolis lizards are provided by \cite{mahler_ingram_revell_losos_2013}.  Our \texttt{R} package is available at \url{https://github.com/WenshaZ/ELPASO}.

\section*{Acknowledgement}
LSTH was supported by the Canada Research Chairs program, the NSERC Discovery Grant RGPIN-2018-05447, and the NSERC Discovery Launch Supplement.
TK was supported by the NSERC Discovery Grant RGPIN/4945-2014.

\bibliography{reference}

\end{document}